\documentclass{aa}  

\usepackage{graphicx}
\usepackage{txfonts}
\usepackage{hyperref}
\usepackage{multirow}

\usepackage{placeins}
\usepackage{stfloats}

\newcommand{\edit}[1]{{#1}}

\begin{document}

\defcitealias{Benitez-Llambay_Pessah_2018ApJ...855L..28B}{BLP18}
\defcitealias{Regaly_2020MNRAS.497.5540R}{Reg\'{a}ly 2020}

   \title{Pebble-driven migration of low-mass planets\\in the 2D regime of pebble accretion}
   \titlerunning{Pebble-driven migration of pebble-accreting planets}


   \author{O. Chrenko
          \inst{1}
          \and
          R. O. Chametla\inst{1}
          \and
          F. S. Masset\inst{2}
          \and
          C. Baruteau\inst{3}
          \and
          M. Bro\v{z}\inst{1}
          }

   \institute{Charles University, Faculty of Mathematics and Physics, Astronomical Institute, V Hole\v{s}ovi\v{c}k\'{a}ch 747/2, 180 00 Prague 8, Czech Republic\\
   \email{chrenko@sirrah.troja.mff.cuni.cz}
   \and
   Instituto de Ciencias F\'{i}sicas, Universidad Nacional Autonoma de M\'{e}xico, Av. Universidad s/n, 62210 Cuernavaca, Mor., M\'{e}xico
   \and
   IRAP, Universit\'{e} de Toulouse, CNRS, UPS, F-31400 Toulouse, France
   }

   \date{Received XXX; accepted YYY}

\abstract
{Pebbles drifting past a disk-embedded low-mass planet develop asymmetries
in their distribution and exert a substantial gravitational torque on the planet, 
thus modifying its migration rate.}
{Our aim is to assess how the distribution of pebbles and the resulting
torque change in the presence of pebble accretion, focusing on its 2D regime.}
{First, we performed 2D high-resolution multi-fluid simulations
with \textsc{Fargo3D} but found that they are impractical for resolving pebble accretion
due to the smoothing of the planetary gravitational potential.
To remove the smoothing and directly trace pebbles accreted by the planet,
we developed a new code, \textsc{Deneb,} which evolves
an ensemble of pebbles, represented by Lagrangian superparticles, in a steady-state
gaseous background.}
{For small and moderate Stokes numbers, $\mathrm{St}\lesssim0.1$, pebble accretion creates
two underdense regions with a front-rear asymmetry with respect to the planet.
The underdensity trailing the planet is more extended. The resulting excess of pebble mass
in front of the planet then makes the pebble torque positive and capable
of outperforming the negative gas torque.
Pebble accretion thus enables outward migration (previously thought to occur mainly for $\mathrm{St}\gtrsim0.1$) in a larger portion of the parameter space.
It occurs for the planet mass $M_{\mathrm{pl}}\lesssim3\,M_{\oplus}$ and for all the Stokes numbers considered in our study,  $\mathrm{St}\in[10^{-2},0.785]$, assuming a pebble-to-gas mass ratio of $Z=0.01$.}
{If some of the observed planets underwent outward pebble-driven migration
during their accretion, the formation sites of their progenitor embryos
could have differed greatly from the usual predictions of planet formation models.
To enable an update of the respective models, we provide a scaling law for the 
pebble torque that can be readily incorporated in N-body simulations.}


   \maketitle
%

\section{Introduction}
\label{sec:intro}

The dynamics of planets born within protoplanetary disks
are governed by their gravitational interaction with the surrounding
disk material. Since gas represents the majority of the disk
mass, it is usually considered to be the sole driver of planetary migration
\citep{Goldreich_Tremaine_1979ApJ...233..857G,Kley_Nelson_2012ARA&A..50..211K,Paardekooper_etal_2023ASPC..534..685P},
acting by means of the resonant \citep{Goldreich_Tremaine_1980ApJ...241..425G,Ward_1986Icar...67..164W,Korycansky_Pollack_1993Icar..102..150K,Ward_1997Icar..126..261W}, dynamical \citep{Paardekooper_2014MNRAS.444.2031P,McNally_etal_2017MNRAS.472.1565M}, magnetohydrodynamic \citep{Baruteau_etal_2011A&A...533A..84B,Uribe_etal_2015ApJ...802...54U,Chametla_etal_2023ApJ...951...81C}, or thermal torques \citep{Lega_etal_2014MNRAS.440..683L,Benitez-Llambay_etal_2015Natur.520...63B,Masset_2017MNRAS.472.4204M,Chrenko_etal_2017A&A...606A.114C}.

\citet[][hereafter BLP18]{Benitez-Llambay_Pessah_2018ApJ...855L..28B}
only recently demonstrated that small solids (dust, pebbles, or boulders), representing
canonically $\sim$$1\%$ of the disk's mass \citep[cf.][]{Lambrechts_Johansen_2014A&A...572A.107L}, can develop 
strong azimuthal asymmetries in their surface density
owing to their interaction with an embedded
low-mass planetary perturber.
The angular momentum exchange during this interaction yields
an additional torque exerted on the planet, hereinafter referred to as the pebble torque\footnote{To avoid confusion, we point out that this term is interchangeable with the term `dust torque' used in \citetalias{Benitez-Llambay_Pessah_2018ApJ...855L..28B}, only we prefer to use 
the `pebble torque' to highlight the range of aerodynamic properties of small solids that we assume.}, which is quite often positive, having an opposite
sign than the gas torque, which is dominated by the negative Lindblad torque 
for a broad range of power-law disks \citep[e.g.][]{Tanaka_etal_2002ApJ...565.1257T,Coleman_Nelson_2014MNRAS.445..479C}.

\citetalias{Benitez-Llambay_Pessah_2018ApJ...855L..28B} performed 2D multi-fluid hydrodynamic simulations to systematically analyse the dependence of the 
pebble torque on the planet mass, $M_{\mathrm{pl}}$,
and the aerodynamic gas-pebble coupling parametrized by the Stokes number,
$\mathrm{St}$ \citep[e.g.][]{Cuzzi_etal_1993Icar..106..102C}.
They find that the scattered pebble flow develops
a downstream overdense filament in front of the planet and
an underdense pebble hole trailing
the planet. The excess of pebbles ahead of the planet and their gravitational
pull are thus responsible for the pebble torque being positive.
The size of the pebble hole is found to increase with 
increasing $\mathrm{St,}$ and the importance of the pebble torque relative
to the gas torque is found to increase with decreasing $M_{\mathrm{pl}}$.
\citetalias{Benitez-Llambay_Pessah_2018ApJ...855L..28B} defined three regimes, namely the gas-dominated regime for which 
the pebble torque is driven mainly by the filament ($\mathrm{St}\lesssim0.25$),
the gravity-dominated regime for which the contribution of the hole
is significant ($\mathrm{St}\gtrsim0.35$), and a transitional regime ($\mathrm{St}\simeq0.25$--$0.35$)
in which an increased concentration of pebbles accumulates at the edge
of the hole and can actually result in a negative pebble torque (due to a mass excess trailing the planet).
Assuming a pebble-to-gas mass ratio\footnote{We note that in high-metallicity disks with low turbulent viscosity, migration of low-mass planets would be further affected by a dynamical corotation torque regulated by the aerodynamic dust feedback, as shown by \citet{Hsieh_Lin_2020MNRAS.497.2425H}.} of $Z=0.01$ and taking an Earth-mass planet as an example, the pebble torque
would overcome the gas torque and revert migration from inwards to outwards
for $\mathrm{St}\gtrsim0.1$, with the exception of 
the transitional regime \citep[$\mathrm{St}\simeq0.25$; see also][]{Guilera_etal_2023ApJ...953...97G}.

When embedded in a sea of pebbles, the planet is also expected to undergo pebble accretion \citep{Ormel_Klahr_2010A&A...520A..43O,Lambrechts_Johansen_2012A&A...544A..32L}. Pebbles become captured by the planet even from large impact parameters (comparable to the planet's Hill radius in the most favourable cases) because the accretion cross-section
is enhanced by the gas drag.
As a pebble becomes deflected by the planet, its kinetic energy is dissipated
by the aerodynamic friction and the pebble can settle onto the planetary surface
\citep{Johansen_Lambrechts_2017AREPS..45..359J,Ormel_2017ASSL..445..197O}.
Since pebble accretion should occur automatically for a planet embedded in a gas-pebble disk, it is definitely worthwhile to examine its influence on 
the pebble distribution generating the pebble torque.

In \citetalias{Benitez-Llambay_Pessah_2018ApJ...855L..28B}, the process of pebble
accretion was neglected. The authors argued that the dynamics of pebbles entering the Hill sphere might not be captured accurately by their model, and they
ignored the disk material located in the inner half of the Hill sphere
in their torque calculations. The first attempt to include pebble accretion
in the context of pebble-driven torque was done by \citet{Regaly_2020MNRAS.497.5540R}. In their multi-fluid model, however, pebble accretion was achieved via an ad hoc reduction of the pebble surface density, with the accretion radius fixed
and the accretion efficiency a free parameter.
In other words, pebble accretion was not achieved
self-consistently by converging pebble trajectories (or streamlines in their case)
at the planet location. One of the difficulties in studying pebble accretion
using global 2D multi-fluid models arises due to the smoothing of the planetary gravitational
potential, as we show in our manuscript \citep[see also][]{Regaly_2020MNRAS.497.5540R}.

In our study, we focused on the 2D limit of pebble accretion in which pebbles are accreted
from a layer that is thinner than the vertical extent of the planet's feeding radius
\citep{Lambrechts_Johansen_2012A&A...544A..32L}.
We mainly relied on a hybrid fluid-particle method in which pebbles
are modelled as Lagrangian superparticles. This approach allowed 
us to remove the smoothing of the planetary potential for pebbles
and track their trajectories very close to the planetary surface,
resolving pebble accretion in a self-consistent manner.
Additionally, the Lagrangian approach captures the effect of 
mutually crossing pebble trajectories, which is not possible
in the fluid approximation. By analysing the pebble torque resulting
from our superparticle model, we show that pebble accretion
can expand the parameter range in which the pebble torque
is positive and supersedes the gas torque.
\section{Methods}

The configuration that we studied corresponds
to a planet with the mass $M_{\mathrm{pl}}$
on a circular orbit at a distance $r_{\mathrm{pl}}=5.2\,\mathrm{au}$
from the central star.
The planet is embedded in a 2D disk of gas and pebbles,
which it perturbs via its gravitational potential.
In this study, the planetary orbit is not evolving,
we only measured the gravitational torque exerted by the
perturbed gas and pebble distributions on the planet.
The reference frame is heliocentric and
corotating with the planet.

\subsection{Pebbles as a fluid}

To establish a firm link to previous works, we performed
multi-fluid simulations in which pebbles are treated as a pressureless
fluid feeling the aerodynamic friction due to its relative
motion with respect to the gas fluid (we neglected the
back-reaction of pebbles on gas).
We used the publicly available
code \textsc{Fargo3D} \citep{Benitez-Llambay_Masset_2016ApJS..223...11B,BL_2019},
which contains the same model ingredients that were
used in \citetalias{Benitez-Llambay_Pessah_2018ApJ...855L..28B}
and we considered a polar mesh with coordinates $(r,\varphi)$.

The gas fluid is treated as a locally isothermal disk with a
constant aspect ratio $h(r)=0.05$. The initial surface density
profile follows
\begin{equation}
    \Sigma_{\mathrm{g}}(r) = (50\,\mathrm{g}\,\mathrm{cm}^{-2})\left(\frac{r}{r_{\mathrm{pl}}}\right)^{-1/2} \, ,
    \label{eq:sigma0}
\end{equation}
the radial gas velocity is initially zero, $v_{r}=0$, and
the rotational gas velocity is
\begin{equation}
    v_{\varphi}(r) = \Omega_{\mathrm{K}}r\sqrt{1-\frac{3}{2}h^{2}} = v_{\mathrm{K}}(1-\eta) \, ,
    \label{eq:vphi0}
\end{equation}
where $\Omega_{\mathrm{K}}$ is the Keplerian frequency,
$v_{\mathrm{K}}=r\Omega_{\mathrm{K}}$ is the Keplerian orbital velocity,
and $\eta$ is the pressure support parameter defined as
\begin{equation}
    \eta = - \frac{1}{2}h^{2}\frac{\partial\log P}{\partial\log r} \, ,
    \label{eq:eta}
\end{equation}
with $P=(hv_{\mathrm{K}})^{2}\Sigma_{\mathrm{g}}$ being the pressure.

The pebble fluid is then initialized via the pebble-to-gas
ratio $Z=0.01$ as
\begin{equation}
  \Sigma_{\mathrm{p}} = Z\Sigma_{\mathrm{g}} \, 
  \label{eq:sigmap0}
,\end{equation}
and the initial velocities follow the expressions\footnote{Please note the corrected last sign in Eq.~(\ref{eq:uphi0}) compared to \citet{Guillot_etal_2014A&A...572A..72G} and \citet{Chrenko_etal_2017A&A...606A.114C}.}
\citep[see][]{Adachi_etal_1976PThPh..56.1756A,Guillot_etal_2014A&A...572A..72G,Chrenko_etal_2017A&A...606A.114C}
\begin{equation}
    u_{\varphi}=v_{\mathrm{K}}-\frac{1}{1+\mathrm{St}^{2}}\left(\eta v_{\mathrm{K}}+\frac{\mathrm{St}}{2}v_{r}\right) \, ,
    \label{eq:uphi0}
\end{equation}
\begin{equation}
    u_{r} = -\frac{2\mathrm{St}}{1+\mathrm{St}^{2}}\left(\eta v_{\mathrm{K}} - \frac{1}{2\mathrm{St}}v_{r}\right) \, .
    \label{eq:ur0}
\end{equation}
The Stokes number $\mathrm{St}$ is
considered a free parameter and expresses the strength of the aerodynamic
drag felt by pebbles:
\begin{equation}
    \vec{a}_{\mathrm{drag}}=\frac{\vec{v}-\vec{u}}{t_{\mathrm{s}}}=\frac{\Omega_{\mathrm{K}}}{\mathrm{St}}\left(\vec{v}-\vec{u}\right) \, ,
    \label{eq:a_drag}
\end{equation}
where $t_{\mathrm{s}}$ is the stopping time.
For the considerations
of our study, it is important to point out
that when the initial conditions of gas and pebbles are combined
together, the radial mass flux of pebbles facilitated by the drag is uniform.
This fact is demonstrated in Appendix~\ref{sec:appA}
and ensures that the mass flux in pebbles drifting past the planet
does not fluctuate in time.

Compared to \citetalias{Benitez-Llambay_Pessah_2018ApJ...855L..28B}, our multi-fluid simulations differ
in the following aspects. The radial grid span is narrower,
covering $0.7$--$1.3\,r_{\mathrm{pl}}$.  
The grid spacing is uniform in all directions,
with the resolution of $N_{r}=2400$ radial rings
and $N_{\varphi}=24576$ azimuthal sectors.
The motivation for this relatively large resolution 
is to maintain the radial spacing equal to the finest level
of the non-uniform grid of \citetalias{Benitez-Llambay_Pessah_2018ApJ...855L..28B}
while doubling the resolution in azimuth to keep 
the grid cells square-shaped at the planet location,
having the size of $2.5\times10^{-4}\,r_{\mathrm{pl}}$.
The Shakura-Sunyaev viscosity parameter \citep{Shakura_Sunyaev_1973A&A....24..337S}
assumed for the gas  fluid is $\alpha=10^{-4}$ \citepalias[compared to $\alpha=3\times10^{-3}$ used in][]{Benitez-Llambay_Pessah_2018ApJ...855L..28B}.

As for the boundary conditions for gas and pebble fluids, we used a
linear extrapolation for the surface densities, a Keplerian extrapolation
for azimuthal velocities, and reflective boundary conditions for radial velocities.
Furthermore, wave-killing zones \citep{deValBorro_etal_2006MNRAS.370..529D}
in which hydrodynamic quantities are damped towards the initial state
were applied at $r<0.77\,r_{\mathrm{pl}}$ and $r>1.18\,r_\mathrm{pl}$.
We point out that we ignored the indirect acceleration term 
related to the reflex motion of the star induced by the disk itself \citep{Crida_etal_2022sf2a.conf..315C}.

In 2D hydrodynamic simulations of planet-disk interactions, it is customary
to consider a smoothed Plummer-type gravitational potential for the planet:
\begin{equation}
    \Phi_{\mathrm{pl}} = -\frac{GM_{\mathrm{pl}}}{\sqrt{d^{2}+\epsilon^{2}}} \, ,
    \label{eq:pot}
\end{equation}
where $d$ is the distance from the planet to a point of interest and
$\epsilon$ is the smoothing length. The purpose of $\epsilon$ is twofold.
First, it is needed to avoid numerical divergence of the potential
term for grid cells overlapping with the planet location. Second,
$\epsilon$ can be fine-tuned
to account for the fact that in reality the planet is not interacting
with a 2D distribution of matter; it is instead interacting with a 3D
disk of non-zero thickness. For isothermal gas disks, it is possible to make
an educated guess of $\epsilon$ so that the gas torque acting on the planet
in the 2D model is similar to the result of 3D simulations
\citep{Muller_etal_2012A&A...541A.123M}. Typically, $\epsilon=0.4$--$0.7H_{\mathrm{g}}$, where
$H_{\mathrm{g}}=hr$ is the pressure scale height.

However, the choice of $\epsilon$ for pebbles is less clear. A comprehensive
comparison of 2D and 3D pebble-driven torque has never been performed;
intuitively, 3D pebble-driven torque will necessarily depend on the level of turbulent
stirring \citep[e.g.][]{Dubrulle_etal_1995Icar..114..237D}. In the two most extreme cases, large pebbles in a laminar disk would settle all the way to the midplane, whereas small pebbles in a turbulent disk might be
completely mixed with the gas, having the same scale height.

For the purpose of our study, it is important to keep backward comparability; 
we thus followed \citetalias{Benitez-Llambay_Pessah_2018ApJ...855L..28B} and set $\epsilon=1\,R_{\mathrm{H}}$ for both gas and pebbles when running
multi-fluid simulations. However, as we show later, this prevents the model from
capturing pebble accretion, and therefore, we also employed
methods that enabled us to relax the value of $\epsilon$ for pebbles.

\subsection{Pebbles as superparticles}
\label{sec:superparticles}

To recover pebble accretion, we examined the limiting 
case in which all pebbles are settled in a thin layer close to the midplane
of the protoplanetary disk. In that case, they are being accreted
in the 2D regime \citep[e.g.][]{Lambrechts_Johansen_2012A&A...544A..32L,Morbidelli_Nesvorny_2012A&A...546A..18M,Morbidelli_etal_2015Icar..258..418M}
and from a physical point of view,
there is no argument for such a pebble disk to evolve in a smoothed planetary
potential; instead one should set $\epsilon=0$.
The dynamics of pebbles within an unsmoothed planetary potential might become very complex, including crossing and repeatedly scattered trajectories
\citep[see e.g.][]{Johansen_Lambrechts_2017AREPS..45..359J}
that the grid-based fluid approximation fails to capture (it mainly captures a local
mean state of multiple pebbles).
Furthermore, it is not clear if the fluid approximation
for pebbles can safely be used with $\epsilon$ smaller
than the minimum size of the grid cells.

For these reasons, we decided
to track individual pebble trajectories using a Lagrangian approach. We developed a new code that reads the gas distribution and velocities
resulting from multi-fluid runs with \textsc{Fargo3D} and evolves an ensemble of 
superparticles within this distribution. 
Since we assumed that the planetary orbit is not evolving and the frame corotates
with the planet, the gas fields obtained by \textsc{Fargo3D} represent a steady state
and are suitable for an ex-post analysis of the pebble evolution.

For future reference, our new hybrid fluid-particle code is named
\textsc{Deneb} (an abbreviation for `dust evolution near embedded bodies').
The trajectories of superparticles are integrated based on their equation of motion
in the heliocentric frame \citep[e.g.][]{Liu_Ormel_2018A&A...615A.138L}:
\begin{equation}
    \frac{\mathrm{d}^{2}\vec{r}}{\mathrm{d}t^{2}} = -GM_{\star}\frac{\vec{r}}{r^{3}}
    + GM_{\mathrm{pl}}\left(\frac{\vec{r}_{\mathrm{pl}}-\vec{r}}{|\vec{r}_{\mathrm{pl}}-\vec{r}|^{3}}-\frac{\vec{r}_{\mathrm{pl}}}{r_{\mathrm{pl}}^{3}}\right)
    + \vec{a}_{\mathrm{drag}} \, ,
    \label{eq:eq_of_motion}
\end{equation}
where $M_{\star}$ is the stellar mass. Of course, the denominator 
of the planet-superparticle acceleration can be again smoothed using
$\epsilon$, if needed. We use the smoothing in superparticle
simulations \edit{only} to enable a comparison with multi-fluid simulations.

Since the underlying gas disk is described on a 2D polar grid,
it is convenient to solve Eq.~(\ref{eq:eq_of_motion}) in polar coordinates
($r$, $\varphi$) and for dynamical variables ($u_{r}$, $l$), where $l=ru_{\varphi}$ is the specific angular momentum of a superparticle. The trajectories
are propagated using the second-order drift-kick-drift scheme of \citet{Mignone_etal_2019ApJS..244...38M} based on the exponential midpoint method.
The whole N-body system is rotated after each drift 
to keep the planet at $\varphi_{\mathrm{pl}}=0$
and to synchronize superparticles with the underlying gas disk.
Gas quantities are inferred at locations of superparticles using
the bilinear interpolation. To convert the number density of superparticles
to the pebble surface density $\Sigma_{\mathrm{p}}$ and map it back to the
polar grid, \edit{the cloud-in-cell method is used}.

Unless specified otherwise, we initialize $N_{\mathrm{sp}}=5\times10^{6}$ superparticles
between $r_{\mathrm{min}}=0.8$
and $r_{\mathrm{max}}=1.2\,r_{\mathrm{pl}}$.
The initial $r(t=0)$ of superparticles
is selected randomly \edit{uniformly between
$0$ and $1$ from the following probability distribution \citep[e.g.][]{Baruteau_etal_2019MNRAS.486..304B}:}
\begin{equation}
    P(r) = \frac{\int_{r_{\mathrm{min}}}^{r}\Sigma_{\mathrm{p}}(r)\mathrm{d}r}{\int_{r_{\mathrm{min}}}^{r_{\mathrm{max}}}\Sigma_{\mathrm{p}}(r)\mathrm{d}r} \, .
    \label{eq:proba}
\end{equation}
The azimuth $\varphi(t=0)$ is sorted out
randomly uniformly between $-\pi$ and $\pi$. Each superparticle is assigned
the same mass $m_{\mathrm{sp}}$
so that the total mass in pebbles 
$M_{\mathrm{p}}=N_{\mathrm{sp}}m_{\mathrm{sp}}$ is equal to $ZM_{\mathrm{g}}$, where $M_{\mathrm{g}}$ is the total gas mass between $r_{\mathrm{min}}$ and $r_{\mathrm{max}}$.
We point out, however, that the superparticles behave
as massless test particles throughout the simulation; the purpose of $m_{\mathrm{sp}}$ is only to reconstruct $\Sigma_{\mathrm{p}}$
and the torque that it exerts on the planet.
The initial velocities of each superparticle are again chosen according to
Eqs.~(\ref{eq:uphi0}) and (\ref{eq:ur0}).

Pebble accretion is achieved in two steps. The first step, as already mentioned,
is to set $\epsilon=0$, which ensures an occurrence of trajectories that converge
and accumulate pebbles at the planet location (see Sect.~\ref{sec:density_vs_accretion}).
The second step is to discard these captured superparticles from the simulation, and we did so
when $d\equiv|\vec{r}_{\mathrm{pl}}-\vec{r}|<3R_{\mathrm{surf}}$, where $R_{\mathrm{surf}}$ is the planetary 
surface radius (obtained from $M_{\mathrm{pl}}$ for the material density $\rho_{\mathrm{pl}}=1.5\,\mathrm{g}\,\mathrm{cm}^{-3}$). The choice of the accretion
distance $3R_{\mathrm{surf}}$
is based on the fact that we do not
properly resolve the formation of the first
planetary atmospheres \citep{Ormel_etal_2015MNRAS.446.1026O,Bethune_Rafikov_2019MNRAS.487.2319B}, which could influence the final stage
of settling of pebbles onto the planet
\edit{\citep[][]{DAngelo_Bodenheimer_2024ApJ...967..124D}}.
However, it is usually safe to assume that pebbles that
manage to enter the Bondi radius eventually become accreted 
\citep[e.g.][]{Popovas_etal_2018MNRAS.479.5136P}.

Drifting superparticles are also discarded from the simulation whenever they 
cross $r_{\mathrm{min}}$.
To preserve the radial mass flux of pebbles towards the planet, new superparticles are introduced in the simulation at $r_{\mathrm{max}}$ to maintain the total number of superparticles in the annulus $r\in[0.92,1]\,r_{\mathrm{max}}$ constant. In our first numerical tests, we found that because the superparticles are revived at a
fixed radial distance, artificial disturbances of $\Sigma_{\mathrm{p}}$
are often produced at the outer boundary because its geometry does not 
reflect the local shape of trajectories --- they often exhibit a kink
either if $\mathrm{St}$ is low and particles cross the outer
gaseous spiral arm or if $\mathrm{St}$ is large and particles perform a small epicyclic
oscillation after passing through the opposition with the planet. 
These effects are summarized in Appendix~\ref{sec:appB}.
The boundary artefacts are undesired because they can spread inwards owing
to the drag-induced drift. To disperse the artefacts, we assigned random noise to the background gas
velocities with an amplitude that is equal to $5\%$ of the respective
velocity component at $r_{\mathrm{max}}$ and decreases linearly to zero as $r$ decreases over the interval $r\in[0.92,1]\,r_{\mathrm{max}}$.

The integration time step in our code depends on the distance from the planet.
When $d>1.75R_{\mathrm{H}}$, we set
$\mathrm{d}t=2\times10^{-3}P_{\mathrm{orb}}$, where $P_{\mathrm{orb}}$ is the orbital timescale of the planet. When $d\in(0.2,1.75]R_{\mathrm{H}}$,
the time step is $\mathrm{d}t=2\times10^{-4}P_{\mathrm{orb}}$.
Finally, we set $t=5\times10^{-6}P_{\mathrm{orb}}$ when $d\leq0.2R_{\mathrm{H}}$. The choice of the time step is
empirical: we reduced the time step at the above-defined ranges
of $d$ until the resulting changes of superparticle trajectories 
became negligible.
We show in Appendix~\ref{sec:appC} that our time step choice
leads to pebble accretion rates matching those from literature.
The \textsc{Deneb} code is further tested 
against an independent fluid-particle code \textsc{Dusty FARGO-ADSG}
\citep{BaruteauZhu2016}
in Appendix~\ref{sec:dusty_comparison}.

\begin{table}[]
    \centering
    \begin{tabular}{lllll}
    \hline \hline
        Set & Code & $\epsilon/R_{\mathrm{H}}$ & $R_{\mathrm{c}}/R_{\mathrm{H}}$ & Pebble removal\\
        \hline
        \texttt{MfaE1} & \textsc{Fargo3D} & $1$ & $0.5$ & N/A \\
        \texttt{MspE1} & \textsc{Deneb} & $1$ & $0.5$ & N/A \\
        \texttt{MspE0} & \textsc{Deneb} & $0$ & $0.1$ & $d<3\,R_{\mathrm{surf}}$ \\
        \hline \hline
        Parameter & \multicolumn{4}{l}{List of values} \\
        \hline
        $M_{\mathrm{pl}}/M_{\oplus}$ & \multicolumn{4}{l}{$0.33$, $0.49$, $0.71$, $1.04$, $1.51$, $2.21$, $3.22$, $4.7$} \\
        $\mathrm{St}$ & \multicolumn{4}{l}{\multirow{2}{6.5cm}{$0.01$, $0.014$, $0.021$, $0.03$, $0.043$, $0.062$, $0.089$, $0.127$, $0.183$, $0.264$, $0.379$, $0.546$, $0.785$}} \\
        & \\
        \hline
    \end{tabular}
    \caption{Overview of simulation
    sets (\emph{top part} of the table)
    and main simulation parameters
    (\emph{bottom part} of the table).
    }
    \label{tab:models}
\end{table}

\subsection{Model overview and diagnostics}

Our main parameters are the planet mass $M_{\mathrm{pl}}$,
the Stokes number of pebbles $\mathrm{St}$, and
the smoothing length of the planetary potential
$\epsilon$ felt by pebbles. We logarithmically sampled
eight and thirteen values from intervals $M_{\mathrm{pl}}\in[0.33,4.7]\,M_{\oplus}$
and $\mathrm{St}\in[0.01,0.785]$, respectively, thus
ensuring an overlap with the slightly more extended parameter space of \citetalias{Benitez-Llambay_Pessah_2018ApJ...855L..28B} (see Table~\ref{tab:models}).

To distinguish between different sets of simulations,
we use abbreviations based on the following conventions:
\texttt{Mfa} stands for `models using the fluid approximation' (i.e. \textsc{Fargo3D} simulations),
\texttt{Msp} stands for `models using superparticles' 
(i.e. \textsc{Deneb} simulations),
and \texttt{E1} or \texttt{E0} indicate whether the
potential smoothing length for pebbles is $\epsilon=1$ or $0\,R_{\mathrm{H}}$.
Our main sets of models are then designated \texttt{MfaE1},
\texttt{MspE1}, and \texttt{MspE0} (see Table~\ref{tab:models}).
When referring to a specific simulation from a given set,
we write, for instance,
\texttt{MfaE1\_M0.3St0.014}, where the suffix gives the respective
values of the planet mass (in Earth masses, $M_{\oplus}$) and the Stokes number.

As \textsc{Fargo3D} calculations are computationally intensive
due to the large domain resolution, and also in order to maintain
backward consistency with \citetalias{Benitez-Llambay_Pessah_2018ApJ...855L..28B}, the simulations \texttt{MfaE1} span 50 orbital periods 
of the planet. On the other hand, our superparticle simulations \texttt{MspE1}
and \texttt{MspE0} are extended
over 200 orbital periods
to ensure that the measured pebble torque is converged (see Sect.~\ref{sec:first_look}).

The torque acting on the planet is evaluated as
\begin{equation}
    \Gamma = \int\limits_{\mathrm{disk}} \Sigma_{i}\left(\vec{r}_{\mathrm{pl}}\times \vec{a}_{\mathrm{g}}\right)\mathrm{d}S \, ,
\end{equation}
where $\Sigma_{i}$ is the torque-generating mass distribution ($\Sigma_{\mathrm{g}}$
for the gas torque or $\Sigma_{\mathrm{p}}$ for the pebble torque),
$\vec{a}_{\mathrm{g}}$ is the specific gravitational acceleration arising from a disk element, and the integral reduces to a discrete sum over the simulated disk patch.
For simulation sets \texttt{MfaE1} and \texttt{MspE1}, we followed
\citetalias{Benitez-Llambay_Pessah_2018ApJ...855L..28B}
and excluded a part of the material located within the Hill sphere
from the torque evaluation. In each grid cell, we multiplied the material mass prior to the
torque computation using the exclusion function
\begin{equation}
  f_{\mathrm{cut}} =
  \begin{cases}
    1 & d > R_{\mathrm{c}} + 0.5R_{\mathrm{H}} \, ,\\
    \sin^{2}{\left[\pi\left(\frac{d-R_{\mathrm{c}}}{R_{\mathrm{H}}}\right)\right]} & d\in[R_{\mathrm{c}},R_{\mathrm{c}}+0.5R_{\mathrm{H}}] \, ,\\
    0 & d < R_{\mathrm{c}} \, ,
  \end{cases}
  \label{eq:cut}
\end{equation}
where $R_{\mathrm{c}}=0.5\,R_{\mathrm{H}}$ is the cutoff radius.
As for our simulation set \texttt{MspE0}, our nominal 
choice is $R_{\mathrm{c}}=0.1\,R_{\mathrm{H}}$, which helps reduce torque oscillations due to the stochasticity
in the delivery of superparticles towards the accretion radius.
However, the choice of $R_{\mathrm{c}}$ is rather arbitrary 
and we thus explore its influence on the torque value in Sect.~\ref{sec:exclusion}.

\begin{figure}
    \centering
    \includegraphics[width=8.8cm]{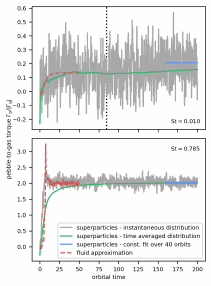}
    \caption{Time evolution of the pebble-to-gas torque ratio, $\Gamma_{\mathrm{p}}/|\Gamma_{\mathrm{g}}|$, as a comparison
    of the fluid and superparticle approximations for 
    the planet mass $M_{\mathrm{pl}}=1.51\,M_{\oplus}$.
    The Stokes numbers are $\mathrm{St}=0.01$ (\emph{top})
    and $0.785$ (\emph{bottom}).
    The planet potential felt by pebbles is smoothed using
    $\epsilon=1\,R_{\mathrm{H}}$.
    The dashed red curve corresponds to simulations
    obtained with the \textsc{Fargo3D} code (set \texttt{MfaE1}).
    For simulations obtained with the \textsc{Deneb} code
    (set \texttt{MspE1},) we distinguish:
    the torque arising from the instantaneous distribution of superparticles
    (grey curve), the torque corresponding 
    to the time-averaged superparticle distribution (green curve),
    and a constant fit of the instantaneous torque (blue line).
    The dotted vertical line is the 
    approximate time it takes pebbles to radially drift
    across the horseshoe region of the planet.}
    \label{fig:torq_vs_hydro}
\end{figure}

\section{Results}

\subsection{First look at the torque measurements}
\label{sec:first_look}

We first explored a typical measurement of the pebble torque
with \textsc{Fargo3D} and \textsc{Deneb}, focusing on
simulation sets \texttt{MfaE1} and \texttt{MspE1}.
Figure~\ref{fig:torq_vs_hydro} shows the
ratio between the pebble torque $\Gamma_{\mathrm{p}}$
and the absolute value of the gas torque $|\Gamma_{\mathrm{g}}|$.
The later is always taken from fluid simulations with \textsc{Fargo3D}
as we assume a non-evolving gas distribution in our superparticle simulations.

\begin{figure}
    \centering
    \includegraphics[width=8.8cm]{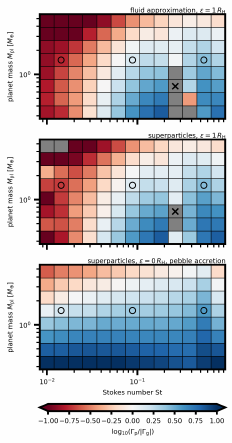}
    \caption{Torque map in the spirit of \citetalias{Benitez-Llambay_Pessah_2018ApJ...855L..28B} showing the logarithm of the pebble-driven torque,
    $\Gamma_{\mathrm{p}}$, scaled by the magnitude of the gas-driven torque,
    $|\Gamma_{\mathrm{g}}|$.
    Grey patches correspond to parameter combinations for which $\Gamma_{\mathrm{p}}<0$. Circles mark the simulations that are discussed
    in detail in Figs.~\ref{fig:dens_cmpr} and \ref{fig:torq_vs_eps_rc},
    while crosses mark the simulation analysed in Figs.~\ref{fig:peculiar_torq} and \ref{fig:peculiar_dens}.
    \emph{Top:} Multi-fluid simulations with \textsc{Fargo3D} (set \texttt{MfaE1}) using the planetary potential smoothing length $\epsilon=1\,R_{\mathrm{H}}$ for both
    gas and pebbles, and the material exclusion radius for the torque calculation $R_{\mathrm{c}}=0.5\,R_{\mathrm{H}}$.
    \emph{Middle:} Simulations with Lagrangian superparticles representing
    pebbles (set \texttt{MspE1}). Parameters are the same as in the \emph{top panel}, except for the time span of 200 orbital periods.
    \emph{Bottom:} Simulations with Lagrangian superparticles 
    avoiding the potential smoothing for pebbles (set \texttt{MspE0}), $R_{\mathrm{c}}=0.1\,R_{\mathrm{H}}$,
    and allowing for pebble accretion.}
    \label{fig:torq_map}
\end{figure}

When measuring $\Gamma_{\mathrm{p}}$ with the \textsc{Deneb} code, 
we tested two approaches. The first approach is to record the torque
arising from the $\Sigma_{\mathrm{p}}(r,\varphi)$ distribution related
to the instantaneous positions of superparticles at the given time $t$.
Since the number of superparticles
within each cell of the 2D mesh fluctuates in time, so does the instantaneous
torque value (grey curve in Fig.~\ref{fig:torq_vs_hydro}). The pebble torque
is then obtained by fitting a constant over a sufficiently long period of time
(blue line in Fig.~\ref{fig:torq_vs_hydro})
defined in our case as the last forty orbital periods of the planet.
The second approach is to record a time-averaged superparticle (and density)
distribution in each cell of the grid and calculate the pebble torque from it (green curve in Fig.~\ref{fig:torq_vs_hydro}). For the purpose of the time averaging,
the superparticle distribution is recorded 50 times per planetary orbit.

In the limit of large Stokes numbers (bottom panel of Fig.~\ref{fig:torq_vs_hydro}; $\mathrm{St}=0.785$),
all torque measurements converge at similar values and we see that there is a very good
agreement between the superparticle and fluid approximations.
However, the situation in the limit of small Stokes numbers (top panel
of Fig.~\ref{fig:torq_vs_hydro}; $\mathrm{St}=0.01$) is different.
We see that while the torque measurements overlap at first
and there seems to be a mock convergence, the torque
value can undergo additional variations
after a timescale roughly equal to the crossing time of the
horseshoe region due to the radial drift of pebbles. Using the horseshoe
half-width $x_{\mathrm{s}}=1.05r_{\mathrm{pl}}\sqrt{M_{\mathrm{pl}}/(hM_{\star})}$ \citep{Jimenez_Masset_2017MNRAS.471.4917J}
and $u_{r}\simeq-2\mathrm{St}\eta v_{\mathrm{K}}$ (Eq.~\ref{eq:ur0}),
we get $t_{\mathrm{cross}}= 2x_{\mathrm{s}}/|u_{r}|\simeq 85 P_{\mathrm{orb}}$
for the top panel of Fig.~\ref{fig:torq_vs_hydro}, as indicated by the vertical dotted line.
In this particular case, one can see that the torque obtained 
in our superparticle simulation starts to slowly increase at $t>t_{\mathrm{cross}}$ and departs from the \textsc{Fargo3D} result.
At the end of the simulation, there is also a mismatch between 
$\Gamma_{\mathrm{p}}$ calculated using the constant fit (blue line) and using the time-averaged
superparticle distribution (green curve). The reason for the difference is that the time-averaged
distribution is plagued by the recordings done at $t<t_{\mathrm{cross}}$ and thus
it takes a relatively large amount of time for it to react to changes of the instantaneous superparticle distribution. 
In the following, all our measurements of $\Gamma_{\mathrm{p}}$
in simulation sets \texttt{MspE1} and \texttt{MspE0}
will be obtained by fitting a constant to the instantaneous torque curve.

\subsection{Parametric study of the pebble torque}

Figure~\ref{fig:torq_map} summarizes the pebble torque 
relative to the gas torque
across the entire parameter space.
Since the same kind of a torque map (showing $\log_{10}(\Gamma_{\mathrm{p}}/|\Gamma_{\mathrm{g}}|)$)
was used in \citetalias{Benitez-Llambay_Pessah_2018ApJ...855L..28B},
one can directly compare a subset of their figure~2 with the top panel
of our Fig.~\ref{fig:torq_map}, corresponding to the simulations
\texttt{MfaE1}.
Unsurprisingly, the comparison yields an excellent agreement.
Minor differences can be found that can be attributed either
to the improved resolution that we used or to the lower 
disk viscosity that we assumed.

Next, \edit{Figure~\ref{fig:torq_map} enables to compare}
the simulation sets \texttt{MfaE1} (top) and \texttt{MspE1} (middle).
Clearly, there is an overall agreement
between the two, which confirms the robustness of the 
superparticle method for measuring $\Gamma_{\mathrm{p}}$.
There are, however, two distinct regions of the parameter
space that exhibit differences. The first region is located
at $\mathrm{St}\lesssim 0.015$. Within this region,
the differences arise as explained in Sect.~\ref{sec:first_look}:
the drift of pebbles is relatively
slow compared to the timescale of the \texttt{MfaE1}
simulations ($50$ orbital periods) and thus the angular momentum
exchange between the planet and pebbles near the horseshoe region
is not necessarily in a steady state.
The second region is located roughly at $\mathrm{St}\in[0.26,0.38]$
and $M_{\mathrm{pl}}\lesssim1.5\,M_{\oplus}$.
It corresponds to the transitional regime defined by
\citetalias{Benitez-Llambay_Pessah_2018ApJ...855L..28B} (see also Sect.~\ref{sec:intro})
and in several cases, it exhibits negative pebble torques (therefore
undefined in the logarithmic scale of Fig.~\ref{fig:torq_map}).
The reason why the superparticle approach predicts different $\Gamma_{\mathrm{p}}$
is that it resolves fine substructures in the pebble flow with
better accuracy (see Appendix~\ref{sec:peculiar}).

Finally, and most importantly, the bottom panel of Fig.~\ref{fig:torq_map}
provides the predicted pebble torque for the simulation set \texttt{MspE0},
that is, when the planetary potential for pebbles is no longer smoothed and
pebble accretion is accounted for.
Substantial changes occur in the map and we summarize them as follows:
\begin{itemize}
    \item The whole parameter
space is now regular and $\Gamma_{\mathrm{p}}>0$ for every explored
combination of $M_{\mathrm{pl}}$ and $\mathrm{St}$.
\item For a fixed planet mass, the dependence of $\Gamma_{\mathrm{p}}$
on $\mathrm{St}$ shows weaker variations than with $\epsilon=1\,R_{\mathrm{H}}$.
\item The pebble torque at $\mathrm{St}\lesssim0.1$ becomes much stronger,
exceeding the gas torque by a factor of several for Earth-sized planets
and by a factor of $10^{1}$ for sub-Earths.
The scaling with the planet mass is the same as at 
$\mathrm{St}\gtrsim0.1$---the lower is the planet mass, the stronger
$\Gamma_{\mathrm{p}}$ becomes relative to the gas-driven torque.
\item At $\mathrm{St}\gtrsim0.1$ (aside from the irregular
region found in top and middle panels of Fig.~\ref{fig:torq_map}), the overall
character of the map remains the same as in simulation sets \texttt{MfaE1}
and \texttt{MspE1}, only the pebble torque magnitude seems to be systematically
shifted to larger values.
\item All these facts suggest that the influence of pebble accretion
in the 2D regime has the largest impact on the pebble torque for low
Stokes numbers (see Sect.~\ref{sec:exclusion} for a confirmation).
This influence is important as it significantly extends
the portion of the parameter space in which
planet migration
can become dominated by the pebble torque in protoplanetary disks.
\end{itemize}

\subsection{Influence of pebble accretion on the torque-generating surface density}
\label{sec:density_vs_accretion}

\begin{figure*}
    \centering
    \includegraphics[width=18cm]{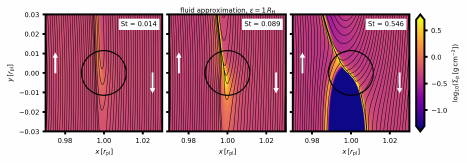}
    \includegraphics[width=18cm]{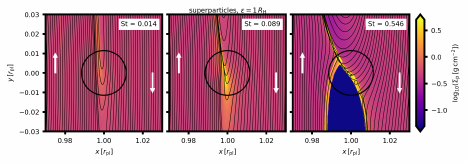}
    \includegraphics[width=18cm]{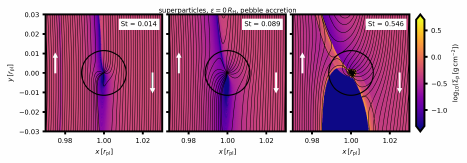}
    \caption{Surface density of pebbles, $\Sigma_{\mathrm{p}}$,
    near a $1.51\,M_{\oplus}$ planet
    for three values of the Stokes number: $\mathrm{St}=0.014$,
    $0.089$, and $0.546$ (\emph{left}, \emph{middle}, and \emph{right columns}, respectively). The black circle shows the extent of the Hill radius,
    and the white arrows show the direction of the shear motion relative to the planet.
    In an inertial frame, the planet would orbit in the $y>0$ direction.
    Radial drift of pebbles occurs in the $x<0$ direction.
    The lower and upper limits of the logarithmic colour palette correspond
    to the initial pebble surface density at the planet location multiplied
    by $10^{-1}$ and $10^{1}$, respectively.
    \emph{Top:} $\Sigma_{\mathrm{p}}$ results directly from \textsc{Fargo3D}
    multi-fluid simulations. The black curves are streamlines of the pebble
    fluid. The planetary potential smoothing for pebbles is $\epsilon_{\mathrm{p}}=1\,R_{\mathrm{H}}$, and the simulation time is $t=50\,P_{\mathrm{orb}}$.
    \emph{Middle:} $\Sigma_{\mathrm{p}}$ calculated from 
    the distribution of superparticles by the cloud-in-cell method and averaged in time. The black curves are trajectories of individual superparticles.
    The potential smoothing for pebbles is $\epsilon_{\mathrm{p}}=1\,R_{\mathrm{H}}$, and the simulation time is $t=200\,P_{\mathrm{orb}}$
    \emph{Bottom:} As in the \emph{middle row} but avoiding potential smoothing for pebbles and allowing for pebble accretion.}
    \label{fig:dens_cmpr}
\end{figure*}

Generally speaking, the pebble torque exists due to a front-rear
asymmetry in the pebble surface density $\Sigma_{\mathrm{p}}$
with respect to the planet.
Therefore, in order to understand what drives the changes of $\Gamma_{\mathrm{p}}$
in Fig.~\ref{fig:torq_map}, it is necessary to examine where
the asymmetry is generated and how it relates to the orbital dynamics of pebbles.
With this in mind, we display $\Sigma_{\mathrm{p}}$ in a close proximity
of a $M_{\mathrm{pl}}=1.51\,M_{\oplus}$ planet in Fig.~\ref{fig:dens_cmpr}.
The rows of Fig.~\ref{fig:dens_cmpr} correspond to simulation sets
\texttt{MfaE1}, \texttt{MspE1}, and \texttt{MspE0}, respectively,
while the columns correspond to $\mathrm{St}=0.014$, $0.089$, and $0.546$, respectively.

The dynamics of pebbles in the first row of Fig.~\ref{fig:dens_cmpr}
(simulations \texttt{MfaE1}) matches 
the findings of \citetalias{Benitez-Llambay_Pessah_2018ApJ...855L..28B}.
When the Stokes number is small and aerodynamic coupling with the 
gas is substantial, one can see that a small part of the pebble streamlines is 
concentrated downstream in front of the planet. This leads to a formation
of an overdense filament responsible for a positive, albeit small, torque. As the Stokes number starts to increase, 
the filament becomes more prominent and starts to extend
also into the region trailing the planet. Pebble streamlines suffer a noticeable deflection when crossing the Hill sphere. Finally, in the limit of large Stokes numbers,
the torque is dominated by a formation of a large-scale underdense hole in the pebble distribution trailing the planet. The filament is present not only in front of the planet, but also behind the planet, encircling the pebble hole. Clearly,
the hole is formed because streamlines preferentially encircle this region, isolating
it from the incoming flux of pebbles. The streamlines encircling the hole exist
due to a combination of planet-induced scattering, which excites epicyclic oscillations
of loosly coupled pebbles passing by the planet, and aerodynamic drag, which damps
the oscillations and eventually enables the pebbles to cross the planet's corotation
by radial drift and encounter the Hill sphere for the second time, approaching 
from the rear \citep[see also][]{Morbidelli_Nesvorny_2012A&A...546A..18M}.

The new result obtained here by examining the close proximity of the planet
and its Hill sphere is that there are no streamlines that would converge at the planet
itself. Hence, the fluid model \texttt{MfaE1} as well as the study of \citetalias{Benitez-Llambay_Pessah_2018ApJ...855L..28B} fail to capture pebble accretion, as we anticipated and as the authors themselves suggested.

The same is true for our superparticle simulations \texttt{MspE1}, in which pebbles
again feel the smoothed planetary potential (see the middle row of Fig.~\ref{fig:dens_cmpr}). The cases of $\mathrm{St}=0.014$ and $0.089$ are hardly
distinguishable from the \texttt{MfaE1} simulations, which serves as a cross-validation
of the two methods and of the \textsc{Fargo3d} and \textsc{Deneb} codes. 
The case of $\mathrm{St}=0.546$ starts to show differences enabled by integrating
the individual trajectories. Some of these trajectories that contribute to the overdense filament perform a loop before exiting the Hill sphere and we also see
that some trajectories cross each other, which is a feature that the fluid approximation
fails to recover since only one fluid velocity vector is defined at each $(r,\varphi)$. The filament exhibits sharp boundaries rather than having a smooth
transition with the background pebble surface density. We emphasize, however,
that the differences in $\Sigma_{\mathrm{p}}$ recognizable in the limit of large Stokes
numbers are not large enough to noticeably modify the torque when comparing
 \texttt{MspE1} and \texttt{MfaE1} simulation sets (see Fig.~\ref{fig:torq_map}).

To appreciate the influence of $\epsilon=0\,R_{\mathrm{H}}$ and the removal of pebbles
by their accretion onto the planet, let us focus on the 
bottom row in Fig.~\ref{fig:dens_cmpr} showing the simulation set \texttt{MspE0}.
For all of displayed cases, we successfully recover trajectories converging
at the planet location \citep[see also][]{Kuwahara_Kurokawa_2020A&A...633A..81K,Visser_etal_2020Icar..33513380V}.
The accretion radius grows with the Stokes number, roughly
as $\propto$$\mathrm{St}^{1/3}$, but we point out that the accretion efficiency in the 2D limit scales inversely, as $\propto$$\mathrm{St}^{-1/3}$ \citep{Lambrechts_Johansen_2012A&A...544A..32L,Morbidelli_etal_2015Icar..258..418M,Liu_Ormel_2018A&A...615A.138L}. The reason is that as the larger pebbles drift
very fast across the planetary orbital radius, a fraction of their flux 
does not even encounter the planet and is protected from pebble accretion.
We verified in Appendix~\ref{sec:appC} that the 2D pebble accretion
rate reproduced by the \textsc{Deneb} code matches previously established 
scaling laws.

For the cases $\mathrm{St}=0.014$ and $0.089$, we notice that the overdense 
filament no longer exists. Instead, there is a new underdense perturbation with a front-rear asymmetry. The perturbation is underdense because a fraction
of pebbles is removed from the flux when encountering the planet.
Pebbles that arrive from the outer disk give rise to the underdense perturbation
trailing the planet after some of them become accreted. 
The accretion of pebbles that arrive from the inner disk,
including some amount of pebbles that 
make a U-turn shortly after encountering the planet for the first time,
results in the formation of the underdense perturbation in front of the planet.
The underdensity trailing the planet is spatially more extended 
and becomes responsible for the large positive torque found in Fig.~\ref{fig:torq_map}.
The extent of the trailing underdensity is large because
with $\epsilon=0\,R_{\mathrm{H}}$, pebble trajectories develop
a protected region (partially disjoint from the pebble flux) similar to the hole
described above even when the Stokes number is low.
The difference between the trailing and leading underdensity is also 
related to a combined effect of the drift and shear motions. The trailing underdensity
forms in a stream of pebbles arriving from the outer disk and as it undergoes inward drift, it crosses the planet's corotation at which the shear direction reverses.
Therefore, the trailing underdensity stagnates in the planet's vicinity.
The leading underdensity, on the other hand, forms below the corotation and thus 
the shear motions and radial drift only spread it away from the planet's vicinity.
Although the leading underdensity reappears 
at $x_{\mathrm{pl}}\simeq0.977\,r_{\mathrm{pl}}$ and catches up with the planet,
it is already too weak and relatively distant to alter the torque.

Regarding the case $\mathrm{St}=0.546$, the main differences 
compared to simulations \texttt{MfaE1} and \texttt{MspE1} are that 
(i) pebbles are accreted preferentially from the outer disk as
the trajectories are angled by the fast radial drift; (ii) 
the filament is less dense as the pebbles no longer escape the potential
well of the planet;
(iii) the filament is wider due to numerous trajectory crossings;
(iv) pebble accretion in the close proximity of the planet proceeds
through a small circumplanetary pebble disk \citep[which was already
observed for $\mathrm{St}\sim1$ in][]{Lambrechts_Johansen_2012A&A...544A..32L}.
However, the large pebble hole is still present and therefore,
the resulting torque remains comparable to simulations
\texttt{MfaE1} and \texttt{MspE1}, only exhibiting a slight boost
of its magnitude (Fig.~\ref{fig:torq_map}, but see also Sect.~\ref{sec:exclusion}).

\begin{figure}
    \centering
    \includegraphics[width=8.8cm]{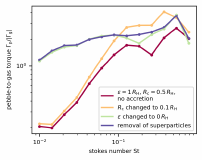}
    \caption{Pebble-to-gas torque as a function of the Stokes number
    for the planet mass $M_{\mathrm{pl}}=1.51\,M_{\oplus}$.
    Individual curves track how the torque changes
    when the key model assumptions are varied one at a time.
    The red curve corresponds to our superparticle simulations \texttt{MspE1}.
    The orange curve is obtained by decreasing the material exclusion
    distance for the torque evaluation to $R_{\mathrm{c}}=0.1\,R_{\mathrm{H}}$.
    The green curve is obtained by decreasing the potential smoothing
    length to zero. The blue curve is obtained by removing superparticles
    below the critical accretion distance and corresponds to the simulation
    set \texttt{MspE0}.
    We point out that the shift from the orange to the green curve
    is caused by the occurrence of pebble-accreting trajectories (such as those
    shown in the bottom row of Fig.~\ref{fig:dens_cmpr}), and therefore,
    the torque-generating asymmetry of the pebble distribution
    is already affected by pebble accretion
    even though superparticles are not yet removed from the simulation (this change
    is shown by the blue curve).
    }
    \label{fig:torq_vs_eps_rc}
\end{figure}

\subsection{Role of material exclusion and superparticle removal for the torque evaluation}
\label{sec:exclusion}

In Fig.~\ref{fig:torq_map}, the shift from the first two panels to the bottom
one involves several changes of model assumptions, mainly the change in the 
potential smoothing length $\epsilon$ for pebbles but also the decrease
in the cutoff radius, $R_{\mathrm{c}}$, for the torque evaluation (Eq.~\ref{eq:cut})
and the removal of superparticles from the simulation at planetocentric
distances $d<3\,R_{\mathrm{surf}}$.
One might therefore rightfully wonder which of these changes is the most important
for the torque map in the presence of pebble accretion. In this section
we try to address this issue by introducing the aforementioned changes one at a time.

Figure~\ref{fig:torq_vs_eps_rc} shows the pebble-to-gas torque as
a function of the Stokes number for the planet mass $M_{\mathrm{pl}}=1.51\,M_{\oplus}$.
To guide the eye, we advise the reader to start from the red curve
corresponding to the simulation set \texttt{MspE1}.
By introducing the decrease in the material exclusion radius from $R_{\mathrm{c}}/R_{\mathrm{H}}=0.5$ to $0.1$, the torque dependence shifts from the red to the
orange curve. Then, by decreasing the potential smoothing
length from $\epsilon/R_{\mathrm{H}}=1$ to $0$, the torque dependence shifts from the orange
to the green curve. Finally, when the removal of superparticles is allowed,
we obtain the blue curve corresponding to the simulation set \texttt{MspE0}.

We can see that as long as $\epsilon/R_{\mathrm{H}}=1$, the reduction
of the material exclusion radius $R_{\mathrm{c}}$ leads to a substantial boost of the pebble torque only at $\mathrm{St}\gtrsim0.07$. To boost the torque
in the regime of smaller Stokes numbers, it is needed to decrease the potential smoothing
to zero, thus enabling the occurrence of pebble-accreting trajectories and the
underdense perturbations near the planet.

Introducing the removal of superparticles from the simulation has actually very little influence on the torque. Such a result is not surprising 
because as long as trajectories that enable pebble accretion are well resolved (which is 
true for the zero smoothing length, i.e. for the green curve in Fig.~\ref{fig:torq_vs_eps_rc}),
they already efficiently remove pebbles from their flux through the Hill sphere
and concentrate them near the planet location.
A difference would only be observed when relaxing the cutoff radius, $R_{\mathrm{c}}$, even
more and thus including fluctuations of the superparticle number density close to the planet
in the torque calculation.

The findings of Fig.~\ref{fig:torq_vs_eps_rc}
align well with our previous assertion that
pebble accretion has the largest influence on the torque
for the smallest considered Stokes
numbers. However, having a correct description of pebble dynamics within
the Hill sphere (at $\epsilon/R_{\mathrm{H}}=0$) is also important for larger
Stokes numbers as it prevents the torque from being overestimated (compare the orange with
the green and blue curves at $\mathrm{St}\gtrsim0.1$).

\subsection{Scaling law for the pebble torque}

\begin{figure}
    \centering
    \includegraphics[width=8.8cm]{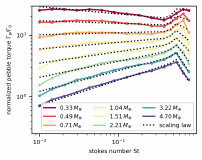}
    \caption{Pebble torque, $\Gamma_{\mathrm{p}}$, normalized
    by $\Gamma_{0}$ (Eq.~\ref{eq:gamma0}) as a function of the Stokes number
    for our superparticle simulations with pebble accretion (set \texttt{MspE0}).
    Coloured curves distinguish
    simulation results for different planetary masses, $M_{\mathrm{pl}}$,
    as specified in the legend. Dotted curves correspond 
    to the empirical scaling law provided in Eqs.~(\ref{eq:torque_1})--(\ref{eq:torque_3})
    and further generalized in Eqs. (\ref{eq:torque_1FIN}) and (\ref{eq:torque_2FIN}).}
    \label{fig:torq_fit}
\end{figure}

\begin{figure}
    \centering
    \includegraphics[width=8.8cm]{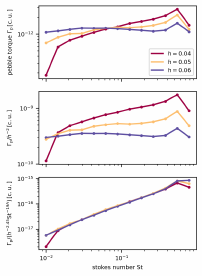}
    \caption{Pebble torque, $\Gamma_{\mathrm{p}}$, acting on a planet with the mass $M_{\mathrm{pl}}=1.51\,M_{\oplus}$
    in gas disks with various aspect ratios, $h$.
    We show the nominal case from the \texttt{MspE0}
    simulations with $h=0.05$ (orange), as well
    as additional calculations for $h=0.04$ (red)
    and $0.06$ (blue). We compare the dependences
    for a non-normalized torque (top), a normalization $\Gamma_{\mathrm{p}}/h^{-2}$ following Eq.~(\ref{eq:gamma0}) (middle), and a 
    normalization, $\Gamma_{\mathrm{p}}/(h^{-2.45}\mathrm{St}^{-19h}),$
    that we found to be intrinsic
    to the pebble torque (bottom). The results of the bottom panel were used to generalize our empirical torque formula in Eqs.~(\ref{eq:torque_1FIN}) and (\ref{eq:torque_2FIN}).}
    \label{fig:torq_scaling_aspect}
\end{figure}

In our simulations with superparticles feeling the unsmoothed 
gravitational potential of the planet, we have already demonstrated in Fig.~\ref{fig:torq_map} that the pebble torque is positive in the entire
parameter space and behaves in a regular fashion. This motivated
us to search for an empirical scaling law that would allow the pebble torque to be easily incorporated in studies with prescribed migration.
These studies involve, but are not limited to, N-body simulations describing
the long-term evolution of disk-embedded planetary
systems \citep[e.g.][]{Izidoro_etal_2017MNRAS.470.1750I,Bitsch_etal_2019A&A...623A..88B,Emsenhuber_etal_2021A&A...656A..70E,Matsumura_etal_2021A&A...650A.116M} or simulations of growth tracks of isolated planets within 1D disks \citep[e.g.][]{Bitsch_etal_2015A&A...582A.112B,Johansen_etal_2019A&A...622A.202J}.

The gas torque is often normalized with respect to \citep[e.g.][]{Kley_Nelson_2012ARA&A..50..211K}
\begin{equation}
    \Gamma_{0} = \left[\Sigma\left(r^{2}\Omega_{\mathrm{K}}\frac{q}{h}\right)^{2}\right]_{\mathrm{pl}} \, ,
    \label{eq:gamma0}
\end{equation}
where $q=M_{\mathrm{pl}}/M_{\star}$ is the planet-to-star mass fraction and the right-hand side is evaluated at the planet's orbital radius. Various components of the gas torque (often depending on the disk model) either directly follow the functional dependence of $\Gamma_{0}$ \citep[e.g.][]{Tanaka_etal_2002ApJ...565.1257T,DAngelo_Lubow_2010ApJ...724..730D}
or remain similar to it within a factor of a few \citep[e.g.][]{Paardekooper_etal_2011MNRAS.410..293P}.
\cite{Guilera_etal_2023ApJ...953...97G}, who was the first to include the 
findings of \citetalias{Benitez-Llambay_Pessah_2018ApJ...855L..28B}
in a global model of planet formation, 
proposed that the pebble torque is likely to be normalized by $\Gamma_{0}$ as well. In the following, however, we show that Eq.~(\ref{eq:gamma0}) does not fully capture the dependence of $\Gamma_{\mathrm{p}}$ on $q$ and $h$. Nevertheless, we also express
the pebble torque as $\Gamma_{\mathrm{p}}/\Gamma_{0}$,
simply because it provides a useful
reference value when comparing to the gas torque.

We find that the pebble torque exerted on low-mass planets
accreting pebbles in the 2D regime can be characterized by the following empirical scaling law
\begin{equation}
    \frac{\Gamma_{\mathrm{p}}}{\Gamma_{0}} = \frac{Z}{0.01}\min\left(1.928\times10^{-3},\xi_{1}(\mathrm{St})\right)q^{\min\left(-0.673,\xi_{2}(\mathrm{St})\right)} \, ,
  \label{eq:torque_1}
\end{equation}
where we defined
\begin{equation}
    \xi_{1}(\mathrm{St}) = 
    \begin{cases}
        2.275\times10^{-2}{\mathrm{St}}^{2.645} & \mathrm{St} \leq 0.45 \, , \\
        9.506\times10^{-7}{\mathrm{St}}^{-12} &
        \mathrm{St} > 0.45 \, , \\
    \end{cases}
    \label{eq:torque_2}
\end{equation}
\begin{equation}
    \xi_{2}(\mathrm{St}) = 
    \begin{cases}
        -0.479+0.2\log{\mathrm{St}} & \mathrm{St} \leq 0.45 \, , \\
        -1.241-0.868\log{\mathrm{St}} & \mathrm{St} > 0.45 \, , \\
    \end{cases}
    \label{eq:torque_3}
\end{equation}
and $\log$ is the natural logarithm.
Figure~\ref{fig:torq_fit} compares the results of our superparticle simulations 
\texttt{MspE0} with the scaling law given by Eqs.~(\ref{eq:torque_1})--(\ref{eq:torque_3})
and shows a good agreement.

Returning to the normalization by $\Gamma_{0}$, we split Eq.~(\ref{eq:gamma0}) into three parts. The part $\propto$$\Sigma_{\mathrm{p}}\Omega_{\mathrm{K}}^{2}r^{4}$ is intrinsic to the pebble torque, 
as we verify in Appendix~\ref{sec:appD}. The part $\propto$$q^{2}$ is not intrinsic 
to the pebble torque: we see in Fig.~\ref{fig:torq_fit} that measurements
of $\Gamma_{\mathrm{p}}/\Gamma_{0}$ for different planet-to-star mass ratios are not aligned, and therefore, the right-hand side of our scaling law (Eq.~\ref{eq:torque_1}) still depends
on $q$.

To investigate the part $\propto$$h^{-2}$, we repeated our superparticle simulations 
\texttt{MspE0} with the planet $M_{\mathrm{pl}}=1.51\,M_{\oplus}$ while changing
the aspect ratio of the gaseous background to $h=0.04$ and $0.06$. A comparison of
the resulting pebble torque is provided in Fig.~\ref{fig:torq_scaling_aspect}. 
From the top panel where we plot $\Gamma_{\mathrm{p}}$ directly, it becomes obvious that the
torque does not simply scale with a power of $h$. 
This is further confirmed by the middle panel, where we plot $\Gamma_{\mathrm{p}}/h^{-2}$,
which is the normalization dictated by Eq.~(\ref{eq:gamma0}) if only $h$ varies.
Clearly, this form of a detrending does not result in an overlap of the measurements.
We numerically found that, to achieve the overlap, the normalization should follow
$\Gamma_{\mathrm{p}}/(h^{-2.45}\mathrm{St}^{-19h})$, as shown in the bottom panel of Fig.~\ref{fig:torq_scaling_aspect}. 
We admit, however, that this normalization
starts to fail for the minimum
and maximum $\mathrm{St}$ used in our study.

The reason why the dependence on $h$ is relatively complex is because the headwind velocity,
to which the dynamics of pebbles is sensitive \citep{Liu_Ormel_2018A&A...615A.138L}, changes with the aspect ratio (see Eq.~\ref{eq:eta}). With the knowledge based on Fig.~\ref{fig:torq_scaling_aspect}, the scaling law for the 
pebble torque can be generalized by replacing Eqs.~(\ref{eq:torque_1}) and (\ref{eq:torque_2})
with
\begin{equation}
    \frac{\Gamma_{\mathrm{p}}}{\Gamma_{0}} = \frac{Z}{0.01}h^{-0.45}\min\left(5\times10^{-4},\xi_{1}(\mathrm{St})\right)q^{\min\left(-0.673,\xi_{2}(\mathrm{St})\right)} \, ,
  \label{eq:torque_1FIN}
\end{equation}
and
\begin{equation}
    \xi_{1}(\mathrm{St}) = 
    \begin{cases}
        5.9\times10^{-3}{\mathrm{St}}^{3.595-19h} & \mathrm{St} \leq 0.45 \, , \\
        2.469\times10^{-7}{\mathrm{St}}^{-11.05-19h} &
        \mathrm{St} > 0.45 \, , \\
    \end{cases}
    \label{eq:torque_2FIN}
\end{equation}
respectively.

To close the section, we point out that 
Eqs.~(\ref{eq:torque_1FIN}), (\ref{eq:torque_2FIN}),
and (\ref{eq:torque_3}) 
were only tested within the parameter range of our analysis and for $h=0.04$--$0.06$.
Whether they can be safely used outside
this range or not remains to be verified. Additionally, a caution is needed
in situations that would violate our model assumptions (see Sect.~\ref{sec:caveats}).

\section{Discussion}
\label{sec:discussion}

\subsection{Pebble torque versus the radial pebble flux}
\label{sec:torq_vs_flux}

\begin{figure}
    \centering
    \includegraphics[width=8.8cm]{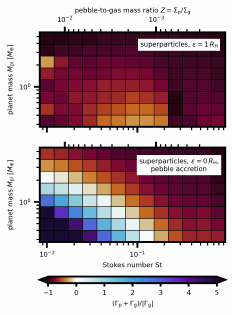}
    \caption{Torque map for a pebble disk 
    characterized by the radial mass flux $\dot{M}_{\mathrm{p}}=10^{-4}\,M_{\oplus}\,\mathrm{yr}^{-1}$. For each Stokes number, $\mathrm{St}$, we 
    obtain $\Sigma_{\mathrm{p}}$ from
    $\dot{M}_{\mathrm{p}}$ (Eq.~\ref{eq:sigma_vs_mdot}), which results in the 
    pebble-to-gas mass ratio, $Z,$ decreasing with increasing $\mathrm{St}$, as shown in the upper horizontal axis. The value of $Z$ is used to rescale the results of Fig.~\ref{fig:torq_map}.
    We show the total torque (exerted by both pebbles and gas) scaled by the gas-driven torque for simulation sets \texttt{MspE1} (\emph{top}) and \texttt{MspE0} (\emph{bottom}).
    In the \emph{top panel}, the total torque is always negative and planet migration is directed inwards.
    In the \emph{bottom panel}, a subset of the parameter space exists for which
    the total torque is positive, resulting in outward migration.}
    \label{fig:discussion}
\end{figure}

Our results imply that the positive pebble torque
can dominate over the negative gas torque
and thus enable outward planet migration in 
a major portion of the parameter space. 
One of the key assumptions that enables the pebble torque to become important,
and one that has not yet been scrutinized 
in our study, is the pebble-to-gas mass ratio, $Z$, which
we set to 0.01,  as in \citetalias{Benitez-Llambay_Pessah_2018ApJ...855L..28B}.
Although $Z=0.01$ is a widely adopted canonical value
characterizing the dust content in protoplanetary disks, it might not be the best parametrization
for pebbles.
The population of pebbles inevitably undergoes depletion by the radial drift while, on the other hand, being replenished by coagulation
from dust-sized particles.
Therefore, it might be more appropriate to parametrize
the pebble disk with the radial mass flux $\dot{M}_{\mathrm{p}}$ rather than assuming a fixed value of $Z$.

\citet{Lambrechts_Johansen_2014A&A...572A.107L}
demonstrated that the typical radial mass flux
in pebbles is $\dot{M}_{\mathrm{p}}\simeq10^{-4}\,M_{\oplus}\,\mathrm{yr}^{-1}$
and the value decreases as the disk evolves. Adopting this finding,
we constructed a new migration map as follows.
First, for each value of $\mathrm{St}$, we 
calculated $\Sigma_{\mathrm{p}}$ from the pebble
mass flux
as outlined in Appendix~\ref{sec:appA} and thus obtained a new estimate of
$Z(\mathrm{St})=\Sigma_{\mathrm{p}}/\Sigma_{\mathrm{g}}$. 
The calculation was done at the radial distance $r_{\mathrm{pl}}$ but, 
as we also show in Appendix~\ref{sec:appA}, the resulting value of $Z(\mathrm{St})$ is radially constant in our simple disk model. For $\dot{M}_{\mathrm{p}}=10^{-4}\,M_{\oplus}\,\mathrm{yr}^{-1}$,
the upper and lower limits in our parameter space
are $Z=1.6\times10^{-2}$ and $3.2\times10^{-4}$
for $\mathrm{St}=0.01$ and $0.785$, respectively.
The value of $Z$ substantially decreases towards larger Stokes numbers because as the radial drift speeds up, the mass flux conservation dictates a lower surface density.
Next, we rescaled each bin of Fig.~\ref{fig:torq_map} by a factor of $Z(\mathrm{St})/0.01$
and thus obtained Fig.~\ref{fig:discussion}.

The top panel of Fig.~\ref{fig:discussion} corresponding to the simulation set
\texttt{MspE1} reveals that outward migration is no longer possible;
the pebble-driven torque can only slow down the rate of inward migration.
When pebble accretion is accounted for, on the other hand, the bottom
panel of Fig.~\ref{fig:discussion} tells us that outward migration 
can still occur in the lower-left corner of the map (i.e. roughly
for $M_{\mathrm{pl}}\lesssim1\,M_{\oplus}$ and $\mathrm{St}\lesssim0.1$).
The rate of outward migration increases with decreasing $M_{\mathrm{pl}}$
and $\mathrm{St}$.
The main takeaway is that the positive torque enabled for $\mathrm{St}\lesssim0.1$
by pebble accretion might represent the main channel of pebble-driven migration,
given that smaller pebbles are depleted less efficiently by the radial drift.

It is interesting to compare this result with \citetalias{Benitez-Llambay_Pessah_2018ApJ...855L..28B}
and \citet{Guilera_etal_2023ApJ...953...97G} who, on contrary, predicted that the pebble torque should
dominate and drive outward migration for $\mathrm{St}\gtrsim0.1$ (the reason for that is immediately visible
when going back to the first two panels of Fig.~\ref{fig:torq_map}). 
Future assessment based on global models of planet
formation is needed to decide whether the predictions of \citet{Guilera_etal_2023ApJ...953...97G}
hold or become replaced with implications of Fig.~\ref{fig:discussion}. Nevertheless, it is clear
that the pebble-to-gas mass ratio has to be treated
with care to correctly determine situations in which
the pebble torque comes into play.

Our discussion only applies
to smooth disks in which the radial pebble mass flux
is not substantially altered by traffic jams or pressure barriers.
However, such dust and pebble concentrations seem
to be ubiquitous in realistic protoplanetary disks
\citep{Andrews_etal_2018ApJ...869L..41A} and
might boost the importance of the pebble torque locally,
due to the sudden increase in $Z$ \citep{Pierens_Raymond_2024A&A...684A.199P}.

\subsection{Possible implications for planet formation}

There are three possible implications for planet formation that
we would like to highlight.
First, it is quite often accepted that gas-driven migration of
Mars-sized and smaller planets is extremely inefficient
given that its speed decreases linearly
with the mass of the migrating body \citep{Tanaka_etal_2002ApJ...565.1257T,Morbidelli_Raymond_2016JGRE..121.1962M}.
On the other hand,
the efficiency of the pebble torque relative to the gas torque increases 
as the planet mass decreases and can reach $\Gamma_{\mathrm{p}}/|\Gamma_{\mathrm{g}}|\sim10^{1}$. Therefore, unless mitigated by the depletion of pebbles
and the decrease in $Z$, the migration rate of these very small
planets might become more substantial than usually thought.

Second, we speculate that convergent planet migration
might occur at special transition radii at which the properties of pebbles
abruptly change. Consider, for instance, the snow line where the water ice
sublimates from inward-drifting pebbles. Outside the snow line, pebbles
are expected to fragment at velocities $\sim$$10\,\mathrm{m}\,\mathrm{s}^{-1}$ \citep{Poppe_etal_2000ApJ...533..454P,Guttler_etal_2010A&A...513A..56G},
whereas they become more fragile as they lose water ice, fragmenting
at $\sim$$1\,\mathrm{m}\,\mathrm{s}^{-1}$ \citep{Gundlach_Blum_2015ApJ...798...34G}.
Since the size of pebbles
scales with the fragmentation velocity squared, it becomes substantially
reduced below the snow line. A sharp transition in the pebble size would result
in smaller Stokes numbers occurring interior to the snow line \citep[e.g.][]{Morbidelli_etal_2015Icar..258..418M,Drazkowska_Alibert_2017A&A...608A..92D,Muller_etal_2021A&A...650A.185M}.
One can envisage a situation in which Earths and sub-Earths 
inside the snow line experience the pebble-driven regime of outward
migration (as in the lower-left corner of the bottom panel in Fig.~\ref{fig:discussion})
but switch to the standard gas-dominated inward migration outside the snow line (by shifting to the orange region of the bottom panel in Fig.~\ref{fig:discussion}).
The snow line could then become a radius of convergent migration,
provided that the radial pebble flux does not
decrease too much due to water ice sublimation \citep{Morbidelli_etal_2015Icar..258..418M}.
Future works should explore the possibility of such an interplay
at various sublimation lines and also at transitions in the turbulent activity of the
disk. For instance, the thermal ionization threshold at $T\simeq900\,\mathrm{K}$ \citep{Desch_Turner_2015ApJ...811..156D} might enable an efficient magnetorotational
instability \citep{Flock_etal_2017ApJ...835..230F}
and truncate pebble sizes at $T\gtrsim900\,\mathrm{K}$, this time due to the increase
in turbulent velocities \citep[e.g.][]{Ueda_etal_2019ApJ...871...10U}.

Last but not least, there are many examples of global
formation models involving pebble accretion
in which planets undergo
only inward migration throughout the lifetime of their natal disk
\citep[e.g.][]{Lambrechts_etal_2019A&A...627A..83L,Johansen_etal_2021SciA....7..444J}.
However, the pebble torque
can facilitate evolutionary stages of outward migration,
having strong implications for  
the initial formation sites of progenitor planetary embryos
or for the material composition inherited by the planets
from the disk.

\subsection{Pebble torque due to accretion of angular momentum}
\label{sec:angmom}

\begin{figure}
    \centering
    \includegraphics[width=8.8cm]{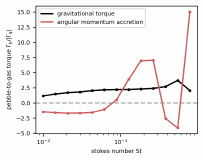}
    \caption{Comparison of the pebble-driven gravitational torque (black) and accretion torque (red; Eq.~\ref{eq:gamma_acc}) for the simulation set \texttt{MspE0} and $M_{\mathrm{pl}}=1.51\,M_{\oplus}$.
    The two torques seem to have a similar level of importance.}
    \label{fig:angmom}
\end{figure}

Throughout our study, we only investigated the pebble
torque related to the gravitational disk-planet interaction.
However, accreted pebbles also carry some amount of angular momentum that might not necessarily match that of the planet. In that case, an additional accretion torque
arises, which amounts to
\begin{equation}
    \Gamma_{\mathrm{acc}} = (\bar{l}_{\mathrm{acc}}-l_{\mathrm{pl}})\dot{M}_{\mathrm{acc}} \, ,
    \label{eq:gamma_acc}
\end{equation}
where we define $\bar{l}_{\mathrm{acc}}$ as the average specific
angular momentum delivered by accreted pebbles, $l_{\mathrm{pl}}$ as the specific angular momentum of the planet, and  $\dot{M}_{\mathrm{acc}}$ as the pebble 
accretion rate of the planet.

In writing Eq.~(\ref{eq:gamma_acc}), we assumed that all of the delivered specific angular momentum modifies the orbital momentum of the planet.
This is a strong assumption made mainly for convenience and in order to highlight a limit scenario. In reality, the accreted angular momentum might as well modify the spin
of the planet \citep{Visser_etal_2020Icar..33513380V,Takaoka_etal_2023A&A...674A.193T,Yzer_etal_2023A&A...678A..37Y}. However, what fraction contributes to the spin and what to the orbital momentum
remains uncertain.

Fig.~\ref{fig:angmom} compares the accretion torque with
the gravitational pebble torque for simulations \texttt{MspE0}, focusing on
the planet mass $M_{\mathrm{pl}}=1.51\,M_{\oplus}$.
Clearly, $\Gamma_{\mathrm{acc}}$ is substantial. 
It can negate the effect of $\Gamma_{\mathrm{p}}$ for $\mathrm{St}\lesssim0.05$
and $\mathrm{St}\simeq0.35$--$0.55$,
or strengthen it for the remaining considered Stokes numbers.
In the presence of $\Gamma_{\mathrm{acc}}$,
the overall picture presented throughout our paper
would change yet again.
Therefore, we will direct our future work to investigate
the influence of the accretion torque in greater detail,
focusing on the final stages of pebble inspiralling 
towards the planet, which requires the primordial atmosphere
to be taken into account (Sect.~\ref{sec:caveats}).

\subsection{Pitfalls of including pebble accretion in 2D multi-fluid simulations}

In recent years, several studies have attempted to include pebble accretion
in 2D hydrodynamic simulations with multiple fluids \citep[e.g.][]{Chrenko_etal_2017A&A...606A.114C,Regaly_2020MNRAS.497.5540R,Pierens_2023MNRAS.520.3286P}. Although different in details, these studies generally
define a pebble accretion radius that is typically a large fraction of the Hill radius and then they reduce the surface density of the pebble fluid
within this whole radius to match the desired accretion rate of the planet.
However, due to the 2D limitation, these studies also employ
the smoothing of the planetary gravitational potential for the pebble fluid.
In the light of our results, such an approach is dangerous. Consider the first
row in Fig.~\ref{fig:dens_cmpr} corresponding to multi-fluid simulations
with the potential smoothing and imagine that the pebble density is reduced
at each time step within a sink area similar to the Hill sphere. 
This reduced density would be advected back into the disk because 
there are no streamlines converging at the planet location.
There is no guarantee that such a change in the pebble distribution
yields a correct torque (any spurious artefact in the front-rear asymmetry of the pebble distribution would change the torque)
and therefore, we warn authors against combining
the material sink approach with the 2D potential smoothing for pebbles.

\subsection{Caveats}
\label{sec:caveats}

In this study, we neglected turbulent dust diffusion
and effects of 3D dynamics.
This should not be a major obstacle for our choice of 
$\alpha=10^{-4}$ (see, for instance, Appendix~\ref{sec:appC} where we show that
the accretion efficiency in a 3D disk would remain similar) but a direct
future validation is desirable. We plan to continue our work by extending
the \textsc{Deneb} code to 3D. Differences might arise, for example, due to
the interaction of pebbles with the recycling flow of low-mass planets \citep{Ormel_etal_2015MNRAS.447.3512O,Kuwahara_Kurokawa_2020A&A...633A..81K}, which has no 2D counterpart, or with their primordial atmosphere.

We should also mention that simulations obtained with the \textsc{Deneb} code 
require that the gas perturbed by the planet
is in a steady state and the planet remains deeply embedded. However, reaching a true steady state might be difficult in low-viscosity disks
because, given enough time, even low-mass non-migrating planets will eventually carve a small dip around their orbit \citep[e.g.][]{Duffell_MacFadyen_2013ApJ...769...41D,Ataiee_etal_2018A&A...615A.110A}.

Furthermore, we assumed that the Stokes number of pebbles is a constant parameter.
It might be more appropriate to fix the physical size of pebbles
and calculate the Stokes number self-consistently based on the background
gas properties \citep{Epstein_1924PhRv...23..710E}.
Additionally, realistic protoplanetary disks contain a
broad distribution of pebble and dust sizes, although such a distribution
might have a well-defined maximum size containing the
majority of the solid mass \citep[e.g.][]{Birnstiel_etal_2012A&A...539A.148B}.

We have already mentioned that pebble-accumulating pressure bumps 
occur in observed disks. The pebble torque might considerably
change in such an environment. Pebble distribution
near a bump is not governed by inward drift; it instead
reaches an equilibrium between convergent drift and 
divergent spreading by turbulent diffusion \citep[e.g.][]{Dullemond_etal_2018ApJ...869L..46D}.
If a planet finds itself in the inner part of the bump where
the gas rotation is super-Keplerian and pebbles undergo outward drift,
it is likely that the pebble hole or the dominant underdensity created by pebble
accretion would form ahead of the planet. Perhaps the sign of the pebble torque
would then flip compared to a smooth sub-Keplerian disk.

Finally, the planet was kept on a fixed circular orbit. Pebble torques acting
on migrating planets or planets with excited orbits are yet to be explored.
The orbital excitation might in fact occur naturally because pebble accretion 
can heat up the planet and trigger its eccentricity growth by means
of thermal forces \citep{Benitez-Llambay_etal_2015Natur.520...63B,Eklund_Masset_2017MNRAS.469..206E,Chrenko_etal_2017A&A...606A.114C,Fromentau_Masset_2019MNRAS.485.5035F,Velasco-Romero_etal_2022MNRAS.509.5622V,Chrenko_Chametla_2023MNRAS.524.2705C}.

\section{Conclusions}

We have investigated the torque acting on disk-embedded low-mass
planets due to the gravitational influence of the surrounding 
distribution of pebbles. Our main objective was to assess
how pebble accretion contributes to the azimuthal asymmetry of the pebble 
distribution with respect to the planet and hence to the pebble torque.
We focused on the 2D limit of pebble accretion, and our parameter
space covered planetary masses $M_{\mathrm{pl}}\in[0.33,4.7]\,M_{\oplus}$
and Stokes numbers of pebbles $\mathrm{St}\in[0.01, 0.785]$.

First, we used the \textsc{Fargo3D} code to show
that the smoothing of the planetary gravitational potential,
which is traditionally employed in 2D multi-fluid simulations
\citepalias{Benitez-Llambay_Pessah_2018ApJ...855L..28B,Regaly_2020MNRAS.497.5540R},
prevents pebble accretion from being captured self-consistently.
To circumvent this issue,
we introduced a new code,
\textsc{Deneb,} that approximates pebbles as Lagrangian superparticles
evolving in a steady-state gaseous background.

We find that pebble accretion does modify
the pebble distribution and influence the torque. When pebbles encounter the planet
and some of them are captured by it, the pebble flux after passing through the encounter region is reduced (pebbles are `missing' in the flux).
For Stokes numbers $\mathrm{St}\lesssim0.1$, pebbles from the outer disk first encounter the planet (mostly) from the front,
thus forming an underdense region that trails the planet.
Pebbles that had already drifted below the planet's corotation encounter the planet (mostly) from the rear, creating an underdense region in front of the planet.
The trailing underdense region is dominant because it is efficiently 
shielded from the incoming pebble flux,
and at the same time it stagnates near the planet's corotation, where the shear
motions reverse.

The paucity of pebbles behind the planet keeps the pebble torque positive and
strong, capable of outperforming the gas torque for $M_{\mathrm{pl}}\lesssim3\,M_{\oplus}$. This was previously thought to be possible
for $\mathrm{St}\gtrsim0.1$ \citepalias{Benitez-Llambay_Pessah_2018ApJ...855L..28B}, but when pebble accretion is accounted for, the pebble torque
starts to dominate planet migration even for small Stokes numbers, $\mathrm{St}\lesssim0.1$. This finding is important because in protoplanetary disks, there are many situations when pebbles are likely to have small Stokes numbers, either because the disk is still young or because the growth of pebbles is inhibited by various physical barriers (due to the radial drift, fragmentation, bouncing, etc).

Regarding larger pebbles with $\mathrm{St}\gtrsim0.1$, 
the main influence of pebble accretion is that there is no sudden
change nor reversal in the pebble torque when $\mathrm{St}\simeq0.3$ \citepalias[i.e. the transitional regime defined by][is not present]{Benitez-Llambay_Pessah_2018ApJ...855L..28B}.
Similar to the case of smaller pebbles, outward migration is again possible 
for $M_{\mathrm{pl}}\lesssim3\,M_{\oplus}$.

All of the above-mentioned conclusions were obtained while assuming a constant pebble-to-gas
mass ratio of $Z=0.01$. When the pebble population is instead characterized
by a uniform radial mass flux of $\dot{M}_{\mathrm{p}}=10^{-4}\,M_{\oplus}\,\mathrm{yr}^{-1}$ \citep{Lambrechts_Johansen_2014A&A...572A.107L}, outward migration
is possible for $M_{\mathrm{pl}}\lesssim1\,M_{\oplus}$ and $\mathrm{St}\lesssim0.1$ but seems unlikely for $\mathrm{St}\gtrsim0.1$ (Sect.~\ref{sec:torq_vs_flux}).

We have provided a scaling law for the pebble torque (Eqs.~\ref{eq:torque_1FIN}, \ref{eq:torque_2FIN}, and \ref{eq:torque_3}) that is
valid in the aforementioned range
of parameters and in 2D locally isothermal disks. This scaling law can be used
for a straightforward inclusion of pebble-driven migration in global models
of planet formation with prescribed migration rates.

We have also shown that the accretion of angular momentum carried by pebbles, in addition to the gravitational torque, 
can modify planet migration (Sect.~\ref{sec:angmom})
and deserves to be studied in detail in follow-up studies.




\begin{acknowledgements}
This work was supported by the Czech
Science Foundation (grant 21-23067M), the Charles University
Research Centre program (No. UNCE/24/SCI/005), and the
Ministry of Education, Youth and Sports of the Czech Republic
through the e-INFRA CZ (ID:90254).
\end{acknowledgements}

%
%

\bibliographystyle{aa}
\bibliography{references}

\begin{appendix}

\section{Uniform radial mass flux of pebbles}
\label{sec:appA}

In this appendix we show that the initial state of
the pebble disk in our models exhibits a uniform radial mass flux
$\dot{M}_{\mathrm{p}}=\mathrm{const}$.
The pebble surface density relates to the mass flux as
\begin{equation}
    \Sigma_{\mathrm{p}}=-\frac{\dot{M}_{\mathrm{p}}}{2\pi r u_{r}} \, .
    \label{eq:sigma_vs_mdot}
\end{equation}
The radial velocity of pebbles (Eq.~\ref{eq:ur0})
can be approximated as
\begin{equation}
    u_{r} \simeq - 2\frac{\mathrm{St}}{1+\mathrm{St}^{2}}\eta v_{\mathrm{K}} \, .
\end{equation}

For a gas disk with $h=\mathrm{const}$ in a locally isothermal approximation,
$P$ is a power-law function of $r$ and thus $\eta \propto (r/P){\partial P}/{\partial r}=\mathrm{const}$ (see Eq.~\ref{eq:eta}). In our study, we additionally considered $\mathrm{St}=\mathrm{const}$
(the Stokes number is only varied between individual simulations as a parameter),
which yields $u_{r}\propto r^{-1/2}$.
Putting everything together, it is clear that $\dot{M}_{\mathrm{p}}=\mathrm{const}$ 
is satisfied whenever
\begin{equation}
    \Sigma_{\mathrm{p}}\propto r^{-1/2} \, ,
\end{equation}
which corresponds to our initial conditions specified in Eqs.~(\ref{eq:sigma0})
and (\ref{eq:sigmap0}).

On the other hand, if the pebble disk is initially parametrized by $\dot{M}_{\mathrm{p}}$,
the same line of reasoning tells us that $Z=\Sigma_{\mathrm{p}}/\Sigma_{\mathrm{g}}=\mathrm{const}$ 
in our disk model. This fact is utilized in Sect.~\ref{sec:torq_vs_flux}.

\section{Removing boundary artefacts in the superparticle distribution}
\label{sec:appB}

\begin{figure}
    \centering
    \includegraphics[width=8.8cm]{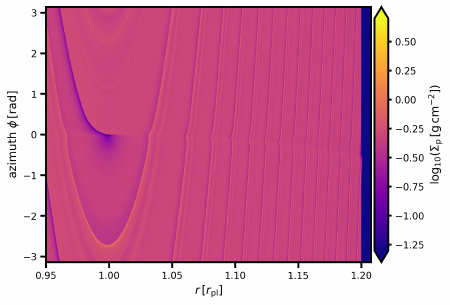}
    \includegraphics[width=8.8cm]{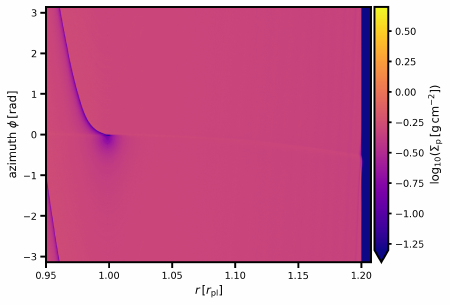}
    \caption{Global view of the time-averaged surface density of pebbles
    in the simulation \texttt{MspE0\_M1.5St0.089}
    without (\emph{top}) and with (\emph{bottom}) the
    addition of random noise to the gas velocities near the outer
    truncation of the superparticle domain.
    The bottom panel correctly recovers the underdense
    perturbation created by pebble accretion and pebble dynamics near the planet
    while avoiding the disturbances spreading from the outer boundary.}
    \label{fig:boundary_artefact}
\end{figure}

In Sect.~\ref{sec:superparticles} we specified that we employed
a narrow patch near the outer boundary of our simulations with superparticles in which we added random noise to the underlying gas velocities. Figure~\ref{fig:boundary_artefact} shows the same simulation
\texttt{MspE0\_M1.5St0.089}
performed without and with the noise (top and bottom panels, respectively).
When the noise is absent, one can see that the time-averaged surface
density develops a dent near the outer boundary $r_{\mathrm{max}}=1.2\,r_{\mathrm{pl}},$ where we introduce new superparticles in the simulation.
The dent is related to the velocity kink that pebbles experience
when crossing the gaseous outer spiral wake.
Consequently, a set of stripes spreads from the dent
by the radial drift and plagues the solution in the domain of interest.
There is a risk that the stripes might cross the planetary orbit near
$\varphi_{\mathrm{pl}}=0$ and affect the torque measurement.
However, it is clear that when the noise zone is considered, the stripes spreading from the outer truncation of the superparticle distribution are 
efficiently erased and the aforementioned problem with the torque
evaluation is avoided.

\section{Verification of accretion rates obtained with the \textsc{Deneb} code}
\label{sec:appC}

\begin{figure}
    \centering
    \includegraphics[width=8.8cm]{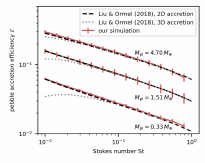}
    \caption{Pebble accretion efficiency, $\mathcal{E,}$ as
    a function of the Stokes number, $\mathrm{St,}$
    for three values of planet mass, $M_{\mathrm{pl}}$,
    across our parameter range. We compare numerical measurements
    obtained with our \textsc{Deneb} code (red points and segments)
    with the scaling laws appropriate for the 2D (dashed black curve)
    and 3D (dotted grey curve) regimes of pebble accretion \citep{Liu_Ormel_2018A&A...615A.138L}. The error bars around the red points correspond to the Poisson
    counting uncertainties.}
    \label{fig:accretion}
\end{figure}

Our code \textsc{Deneb} introduced in Sect.~\ref{sec:superparticles}
has been developed specifically for this study
and it is only appropriate to present its validation tests.
One proof of validity is presented in Fig.~\ref{fig:torq_map}
where the first two panels compare the pebble torques resulting from
\textsc{Fargo3D} simulations
with the outcome of \textsc{Deneb}.
Another proof is provided in Fig.~\ref{fig:dens_cmpr} where
the particle trajectories are indistinguishable from 
pebble fluid streamlines as long as $\mathrm{St}$ is low
and $\epsilon$ is the same.
Here we focus on the measurement of pebble accretion rates
obtained with our new code. More specifically, we followed 
\citet{Liu_Ormel_2018A&A...615A.138L} and 
measured the accretion
efficiency $\mathcal{E}_{\mathrm{2D}}=\dot{M}_{\mathrm{acc}}/\dot{M}_{\mathrm{p}}$,
where $\dot{M}_{\mathrm{acc}}$ is the pebble accretion rate.

\FloatBarrier

The results are presented in Fig.~\ref{fig:accretion}, where we plot
$\mathcal{E}_{\mathrm{2D}}$ as a function of $\mathrm{St}$
for three different planetary masses $M_{\mathrm{pl}}$. The measurement
was performed as follows. We initialized $10^{3}$ particles (or $10^{4}$
particles for the lowest considered planet mass)
at the radial distance $r=r_{\mathrm{pl}}+10R_{\mathrm{H}}$
uniformly along the whole azimuth. We then integrated the particle
trajectories until they either became accreted by the planet
or they crossed $r=r_{\mathrm{pl}}-2R_{\mathrm{H}}$ due to their radial
drift. The efficiency $\mathcal{E}_{2D}$ was computed directly
as the fraction of particles that hit the planet.
As our intention was to compare the resulting $\mathcal{E}_{\mathrm{2D}}$
with \citet{Liu_Ormel_2018A&A...615A.138L}, we azimuthally averaged
the gas velocities $v_{\varphi}$ and set $v_{r}=0$
because \citet{Liu_Ormel_2018A&A...615A.138L} did
not account for perturbations of gas induced by the planet.

The results of Fig.~\ref{fig:accretion} reproduce the scaling law
of \citet{Liu_Ormel_2018A&A...615A.138L} very well (within the margins
of the Poisson counting errors). Furthermore, we plot
the 3D accretion rate $\mathcal{E}_{\mathrm{3D}}$
\citep{Liu_Ormel_2018A&A...615A.138L,Ormel_Liu_2018A&A...615A.178O} to compare it with the 2D approximation.
In the 3D accretion formula, we considered the vertical scale height
of pebbles \citep{Dubrulle_etal_1995Icar..114..237D}
\begin{equation}
    H_{\mathrm{p}} = \sqrt{\frac{\alpha}{\alpha+\mathrm{St}}}H_{\mathrm{g}} \, .
    \label{eq:Hp}
\end{equation}
We can see that since the disk viscosity $\alpha=10^{-4}$ adopted in our
models is low, the resulting turbulent stirring of the pebble disk
would be relatively weak and thus the 3D feeding zone of the 
planets would encapsulate the majority of the vertical height of the pebble layer,
making 2D and 3D accretion rates compatible in the
explored range of Stokes numbers $\mathrm{St}\in[0.01, 0.785]$.
The largest difference is found for $M_{\mathrm{pl}}=0.33\,M_{\oplus}$ and $\mathrm{St}=0.01$, predicting the 2D accretion rate by a factor of $\simeq$$1.8$ larger
than the 3D accretion rate.

\section{Comparison of the \textsc{Deneb} and \textsc{Dusty FARGO-ADSG} codes}
\label{sec:dusty_comparison}

\begin{figure}[!ht]
    \centering
    \includegraphics[width=8.8cm]{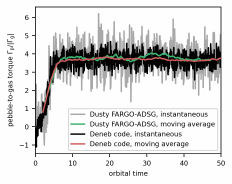}
    \includegraphics[width=8.8cm]{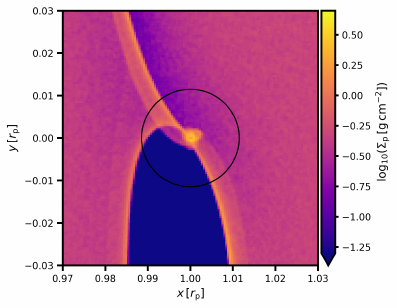}
    \includegraphics[width=8.8cm]{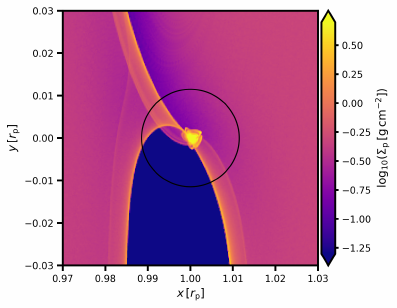}
    \caption{Comparison between the \textsc{Dusty FARGO-ADSG}
    and \textsc{Deneb} codes for the simulation \texttt{MspE0\_M1.5St0.546}.
    We show the time evolution of the pebble-to-gas torque, $\Gamma_{\mathrm{p}}/|\Gamma_{\mathrm{g}}|,$ over 50 orbital periods (\emph{top})
    and the time-averaged pebble surface density, $\Sigma_{\mathrm{p}}$,
    after simulating 50 orbital periods with \textsc{Dusty FARGO-ADSG}
    (\emph{middle}) and 200 orbital periods with \textsc{Deneb} (\emph{bottom}).
    }
    \label{fig:dusty_comparison}
\end{figure}

As our final validation test of the \textsc{Deneb} code, we performed
a  comparison with \textsc{\href{https://github.com/charango/dustyfargoadsg}{Dusty FARGO-ADSG}} \citep{BaruteauZhu2016},
which is a 2D hybrid fluid-particle code as well.
However, \textsc{Dusty FARGO-ADSG} evolves the gas and pebbles
simultaneously, with the time-step size being controlled by the
Courant-Friedrichs-Lewy condition appropriate for the studied system.
Additional differences of our \textsc{Dusty FARGO-ADSG} runs include:
\begin{itemize}
    \item A semi-implicit first-order Euler orbital integrator \citep[similar to][]{Zhu_etal_2014ApJ...785..122Z} is used to evolve superparticles.
    \item Interpolation back and forth between the grid-based quantities
    and their counterparts
    at superparticle locations is achieved using a triangular-shaped cloud
    method.
    \item The grid resolution is lower, $N_{r}\times N_{\varphi} = 1200 \times 12288$.
    \item The number of superparticles is lower, $N_{\mathrm{sp}}=10^{6}$.
    \item The replenishment of superparticles at $r_{\mathrm{max}}$
    is slightly different; each superparticle that becomes accreted or drifts below
    $r_{\mathrm{min}}$ becomes immediately revived at $r_{\mathrm{max}}$.
    \item The distance from the planet below which superparticles are accreted is larger, equal to $0.03\,R_{\mathrm{H}}$.
\end{itemize}

Figure~\ref{fig:dusty_comparison} compares the torque evolution 
and time-averaged pebble surface density between \textsc{Dusty FARGO-ADSG}
and \textsc{Deneb} for the simulation \texttt{MspE0\_M1.5St0.546}
(corresponding also to the last panel of Fig.~\ref{fig:dens_cmpr}).
The match between the torque, as well as between the main features of the 
pebble distribution, is satisfactory and tells us that:
\begin{itemize}
    \item The results obtained with \textsc{Deneb} are robust against
    specific choices of methods listed above (e.g. the orbital integrator
    or the interpolation method).
   \item Our assumption of the gas distribution being non-evolving in \textsc{Deneb}
is valid for the physical model that we have considered.
   \item High resolution and high number of superparticles is beneficial to reduce the 
noise of the torque measurements and resolve fine details in the pebble distribution.
\end{itemize}

\section{Peculiar pebble traps in the presence of the smoothed planet potential}
\label{sec:peculiar}

\begin{figure}
    \centering
    \includegraphics[width=8.8cm]{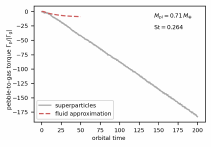}
    \caption{Temporal evolution of the pebble-to-gas torque for simulations
    \texttt{MfaE1\_M0.7St0.264} (dashed red curve) and \texttt{MspE1\_M0.7St0.264}
    (grey curve). The linear decrease in the pebble torque is directly
    related to the accumulation of pebbles in a filament trailing the planet
    (see Fig.~\ref{fig:peculiar_dens}).}
    \label{fig:peculiar_torq}
\end{figure}

\begin{figure*}[b]
    \centering
    \includegraphics[width=18cm]{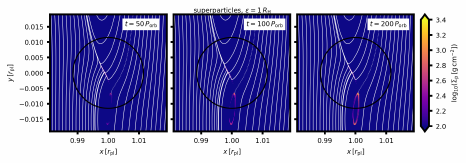}
    \caption{Evolution of the time-averaged surface density,
    $\Sigma_{\mathrm{p}}$, in the simulation
    \texttt{MspE1\_M0.7St0.264}. The time is given by a label
    in each panel. The colour palette is saturated in order to 
    highlight how the overdensity of the filament encircling the pebble hole
    increases in time as more and more pebbles become trapped. A representative trajectory entering the overdense region is shown with a dotted curve and
    then truncated in order to keep the filament visible.
    The filament generates a strong negative torque (see Fig.~\ref{fig:peculiar_torq}).}
    \label{fig:peculiar_dens}
\end{figure*}

Having a deeper understanding of the typical sub-structures in the pebble
distribution that drive the torque (Sect.~\ref{sec:density_vs_accretion}),
\edit{here we examine} one of the cases
corresponding to the transitional regime \citepalias[defined in][]{Benitez-Llambay_Pessah_2018ApJ...855L..28B} for which the simulation sets \texttt{MfaE1} and \texttt{MspE1} disagreed.
\edit{The simulation parameters were} $M_{\mathrm{pl}}=0.71\,M_{\oplus}$ and $\mathrm{St}=0.264$
(simulations marked with crosses in Fig.~\ref{fig:torq_map}).

Figure~\ref{fig:peculiar_torq} shows the pebble-to-gas torque recorded in the respective
simulations. We can see that although the pebble torque is negative in both cases,
it does show signs of convergence in the fluid approximation while 
it steadily continues to decrease in the superparticle approximation.
At the end of the superparticle simulation, the pebble torque exceeds the gas
torque by more than two orders of magnitude, which would of course trigger
fast inward migration of the planet.

The persistent decrease in the pebble torque shown in Fig.~\ref{fig:peculiar_torq}
suggests that there must be an overdensity of pebbles trailing the planet
and that this overdensity traps more and more pebbles as the simulation progresses.
Furthermore, since the decrease in the torque only appears for the superparticle simulation,
we expect the overdensity to be related to structures at the resolution limit of the underlying grid. We point out that \citetalias{Benitez-Llambay_Pessah_2018ApJ...855L..28B} already reported that a very large resolution is needed to properly capture
the torque behaviour in the transitional regime.

Indeed, this is confirmed in Fig.~\ref{fig:peculiar_dens}, which shows the time evolution of the pebble surface density in the superparticle simulation.
We can see that the density of the filament encircling the pebble hole grows in time,
explaining why the torque becomes ever more negative.
The peak density in the filament at the end of the simulation is huge,
reaching $\Sigma_{\mathrm{p}}\simeq2500\,\mathrm{g}\,\mathrm{cm}^{-2}$
and $Z\simeq50$. The filament is very thin compared to the cell sizes
of the underlying polar grid, which is why the fluid approximation
leads to an underestimation of the local density accumulation.

Dynamical effects that can trap pebbles are of increased importance in
the field of planet formation because they can promote further coagulation-fragmentation cascade in the local size distribution,
gas-dust instabilities, or even a collapse into larger objects.
Nevertheless, it is questionable whether the interplay described above is realistic.
First, it occurs when pebbles feel the smoothed planetary potential, which, as we argued, is undesired for pebble accretion.
Second, it completely breaks our model assumptions. Since the solid-to-gas ratio
increases by several orders of magnitude, the aerodynamic back-reaction of pebbles
on the gas cannot be ignored anymore.
Similarly, because the torque on the planet is strong,
the whole interplay should be studied with a moving planet, not a fixed one.
Finally, it is likely that even a small level of turbulent diffusion would
reduce the amount of trapped pebbles, given the sharp density gradient
and the small spatial scale on which it occurs across the filament.
We leave these investigations for future work.

\section{Pebble torque normalization}
\label{sec:appD}

\begin{figure}
    \centering
    \includegraphics[width=8.8cm]{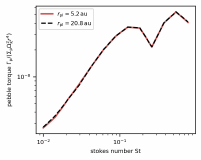}
    \caption{Pebble torque normalized as $\Gamma_{\mathrm{p}}/(\Sigma_{\mathrm{p}}\Omega_{\mathrm{K}}^{2}r^{4})$ for two different radial distances of the planet.
    The fluid approximation for pebbles
    is used, corresponding to the model \texttt{MfaE1}.
    The overlap of the measurements suggests that this normalization is intrinsic to
    the pebble torque.}
    \label{fig:torq_scaling}
\end{figure}

In this appendix we explore whether
the pebble torque can be intrinsically normalized 
with a factor $\propto$$\Sigma_{\mathrm{p}}\Omega_{\mathrm{K}}^{2}r^{4}$,
which represents a part of the quantity 
$\Gamma_{0}$ introduced in Eq.~(\ref{eq:gamma0}).
To verify that,
we repeated the simulations \texttt{MfaE1} with a $M_{\mathrm{pl}}=1.51\,M_{\oplus}$
planet but we shifted the planet to $r_{\mathrm{pl}}=20.8\,\mathrm{au}$.
Then we compared the resulting torque normalized as $\Gamma_{\mathrm{p}}/(\Sigma_{\mathrm{p}}\Omega_{\mathrm{K}}^{2}r^{4})$ with 
our nominal measurement at $r_{\mathrm{pl}}=5.2\,\mathrm{au}$.
Figure~\ref{fig:torq_scaling} depicts this comparison for our range of Stokes numbers
and validates the normalization, at least
for the model assumptions adopted in our study.

\end{appendix}

\end{document}